\documentclass[aps,pra,amsmath,amssymb,floatfix,twocolumn,amsmath,superscriptaddress,twocolumn,nofootinbib,tighten,letterpaper]{revtex4-2}

\usepackage{tikz}
\usepackage{lmodern}

\usepackage{amsbsy}
\usepackage{amsthm}
\usepackage[autostyle]{csquotes}
\usepackage{setspace}
\usepackage{svg}
\usepackage{bm}
\usepackage{placeins}

\usepackage{tabularray}
\definecolor{RowColor}{rgb}{0.88,1,0.9}

\newcommand{\bra}[1]{\left< #1 \right|}
\newcommand{\ket}[1]{\left| #1 \right>}
\usepackage{textgreek}

\usepackage{multirow}
\usepackage{subfigure}
\usepackage{color}
\usepackage{mathrsfs}
\usepackage{hyperref}
\usepackage[normalem]{ulem}
\usepackage{bm}

\usepackage{amssymb}   
\usepackage{amsmath}
\renewcommand\vec[1]{\ensuremath\boldsymbol{#1}} 

\usepackage{amsfonts, relsize}
\usepackage{graphics}
\usepackage{graphicx}
\usepackage{subfigure}
\usepackage{hyperref}
\usepackage{color}
\usepackage{comment}

\begin{document}
\title{Non-unitary time dynamics of topological modes in open planar quantum systems}

\author{Saakshi Porwal}
\affiliation{Department of Physics, Indian Institute of Science, Bangalore 560012, India}

\author{Bitan Roy}
\affiliation{Department of Physics, Lehigh University, Bethlehem, Pennsylvania, 18015, USA}

\date{\today}

\begin{abstract}
Nontrivial topological invariant of bulk electronic wavefunctions in two-dimensional quantum crystals leaves its footprints on the edge, dislocation, and corner modes. Here we investigate non-unitary time dynamics of these topological modes in square lattice-based open quantum systems in which the time-dependent Hamiltonian smoothly interpolates between topologically distinct insulators across band gap closing quantum critical points. The temporal dynamics of these modes is described by a Lindblad equation in which the instantaneous Hamiltonian plays the role of the Lindblad operator, thereby allowing the environment to couple with the system through the energy channels (weak measurement protocol). We show that in the presence of such a real time ramp, the survival probability of these modes decreases (increases) in short (long) time scale where the dephasing (quantum Zeno) effect dominates with the increasing amplitude of the system-to-environment coupling, for both slow and fast ramps from a topological to a normal insulating state. For a reverse course of the time evolution, the revival or condensation probability of nucleating such topological modes, otherwise absent in the initial system, increases for stronger system-to-environment coupling. This phenomenon can be attributed to the strong decoherence of the initial mixed state among all the energy eigenstates of the final Hamiltonian which also includes the topological modes, causing their enhanced condensation probability. Our findings can be germane to real open topological materials with time-tunable band gap, and should be applicable to open topological crystals of arbitrary dimension and belonging to any symmetry class.  
\end{abstract}

\maketitle

\section{Introduction}

\begin{figure}[t!]
\includegraphics[width=1.00\linewidth]{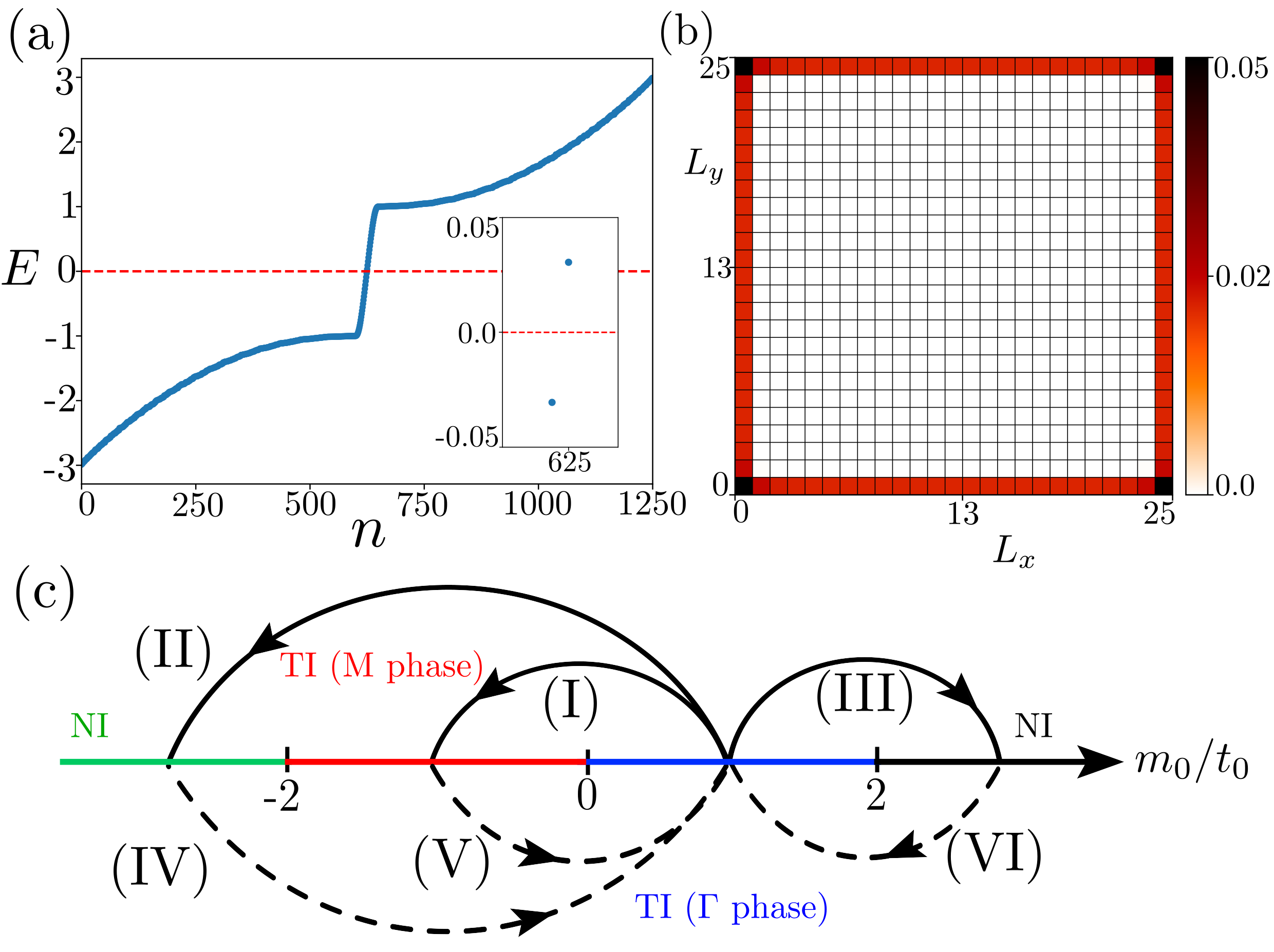}
\caption{(a) Energy ($E$) spectra of the static Qi-Wu-Zhang model Hamiltonian [Eq.~\eqref{eq:hamiltonian}] on a square lattice with open boundary conditions in the $x$ and $y$ directions for $t_1=t_0=m_0=1$, yielding the $\Gamma$ phase with the band inversion at the $\Gamma$ point of the Brillouin zone. Here, $n$ is the eigenvalue index. Inset shows two closest to zero-energy modes. (b) The local density of states of those two near zero-energy modes is highly localized near the edges of the square lattice. (c) Phase diagram of the Qi-Wu-Zhang model. The real time ramp out of (into) the $\Gamma$ phase to (from) other insulating phases are shown by solid (dashed) arrows, labeled by Roman numerals. The corresponding dynamic melting (condensation) of the edge modes are shown in Figs.~\ref{fig:EdgeMeltingProb}-\ref{fig:EdgeMeltingLDOS3} (Fig.~\ref{fig:EdgeCondensation}). Here, TI (NI) stands for topological (normal) insulator.  
}~\label{fig:SetupEdge}
\end{figure}

Nature fosters a variety of topological phases of matter that can be classified into two broad categories, gapped and gapless or nodal. Both classes of topological phases can be found for charged~\cite{TI:1, TI:2, TI:3, TI:4, TI:5, TI:6, TI:7, TI:8, TI:9, TI:10, TI:11, TI:12, TI:13} and neutral Majorana fermions~\cite{TSC:1, TSC:2, TSC:3, TSC:4, TSC:5, TSC:6}, with the later ones corresponding to topological superconductors. Irrespective of these details, all topological phases sustain gapless modes at the boundaries of the underlying host crystal, where the translational symmetry is broken. These modes are protected by the topological invariant of the bulk fermionic wave-functions, capturing their non-trivial geometry, a phenomenon known as the bulk-boundary correspondence, with our focus being restricted on two-dimensional (2D) topological insulators (TIs) in this work. Typically, a $d$-dimensional topological crystal features gapless modes on $(d-1)$-dimensional boundary. In that case, the boundary modes are characterized by the co-dimension $d_c=d-(d-1)=1$ and we realize a first-order topological phase. Therefore, a 2D first-order TI, such as the quantum anomalous Hall insulator (QAHI), supports one-dimensional topological edge modes, see Fig.~\ref{fig:SetupEdge}.

\begin{figure}[t!]
\includegraphics[width=1.00\linewidth]{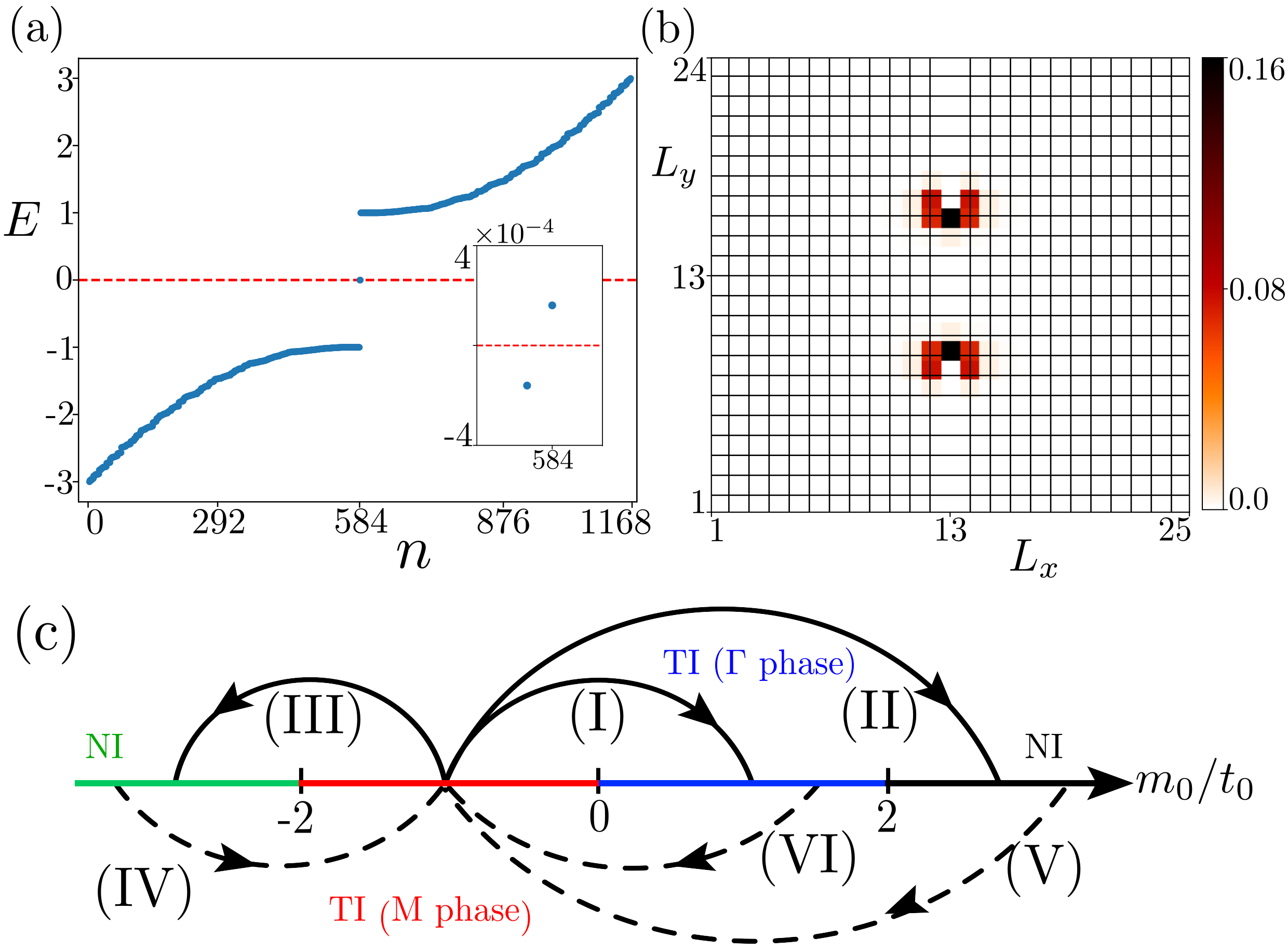}
\caption{(a) Energy ($E$) spectra of the static Qi-Wu-Zhang model Hamiltonian [Eq.~\eqref{eq:hamiltonian}] on a square lattice with periodic boundary conditions in the $x$ and $y$ directions for $t_1=t_0=-m_0=1$, yielding the translationally active ${\rm M}$ phase with the band inversion at the ${\rm M}$ point of the Brillouin zone in the presence of an edge dislocation-antidislocation pair. Here, $n$ is the eigenvalue index. Inset shows two closest to zero-energy modes. (b) The local density of states of those two near zero-energy modes is highly localized around the cores of dislocation and antidislocation. (c) Phase diagram of the Qi-Wu-Zhang model. The real time ramp out of (into) the ${\rm M}$ phase to (from) translationally inert topological $\Gamma$ phase and normal insulating phases are shown by solid (dashed) arrows, labeled by Roman numerals. The corresponding dynamic melting (condensation) of the dislocation modes is shown in Figs.~\ref{fig:DislocationMeltingProb}-\ref{fig:DislocationMeltingLDOS3} (Fig.~\ref{fig:DislocationCondensation}). Here, TI (NI) stands for topological (normal) insulator.  
}~\label{fig:SetupDislocation}
\end{figure}

The underlying mechanism responsible behind topological insulation is the band inversion that can take place either at the center of the Brillouin zone (BZ) known as the $\Gamma$ point or at one of the finite time-reversal-invariant-momentum (TRIM) points (${\bf K}_{\rm inv}$) therein. Finite TRIM points exist in the BZ due to the translational symmetry of the underlying lattice~\cite{defect:0}. A TI with a finite ${\bf K}_{\rm inv}$ is thus also named \emph{translationally active} TI. Besides harboring boundary modes, they also respond nontrivially in the presence of underlying lattice defects, such as dislocations, realized in the interior of the lattice by breaking the translational symmetry locally around the core of such crystal defects. Dislocation lattice defects are characterized by the Burgers vector ${\bf b}$. In particular, when ${\bf K}_{\rm inv} \cdot {\bf b}=\pi$ (modulo $2\pi$), midgap zero-energy modes get pinned around the core of dislocation lattice defects. The ${\bf K} \cdot {\bf b}$ rule is otherwise operative in static, dynamic or Floquet, and non-Hermitian topological crystals~\cite{defect:1, defect:2, defect:3, defect:4, defect:5, defect:6, defect:7, defect:8, defect:9, defect:10, defect:11, defect:12, defect:13, defect:14, defect:15}. An example of such dislocation modes in a QAHI with the band inversion at the ${\rm M}=(\pi,\pi)/a$ point of the BZ is shown in Fig.~\ref{fig:SetupDislocation}, where $a$ is the lattice spacing of the underlying square lattice. On the other hand, when the band inversion occurs at the $\Gamma=(0,0)$ point of the BZ, ${\bf K}_{\rm inv} \cdot {\bf b}=0$, and the resulting TI does not host any midgap zero-energy dislocation modes. 

\begin{figure}[t!]
\includegraphics[width=1.00\linewidth]{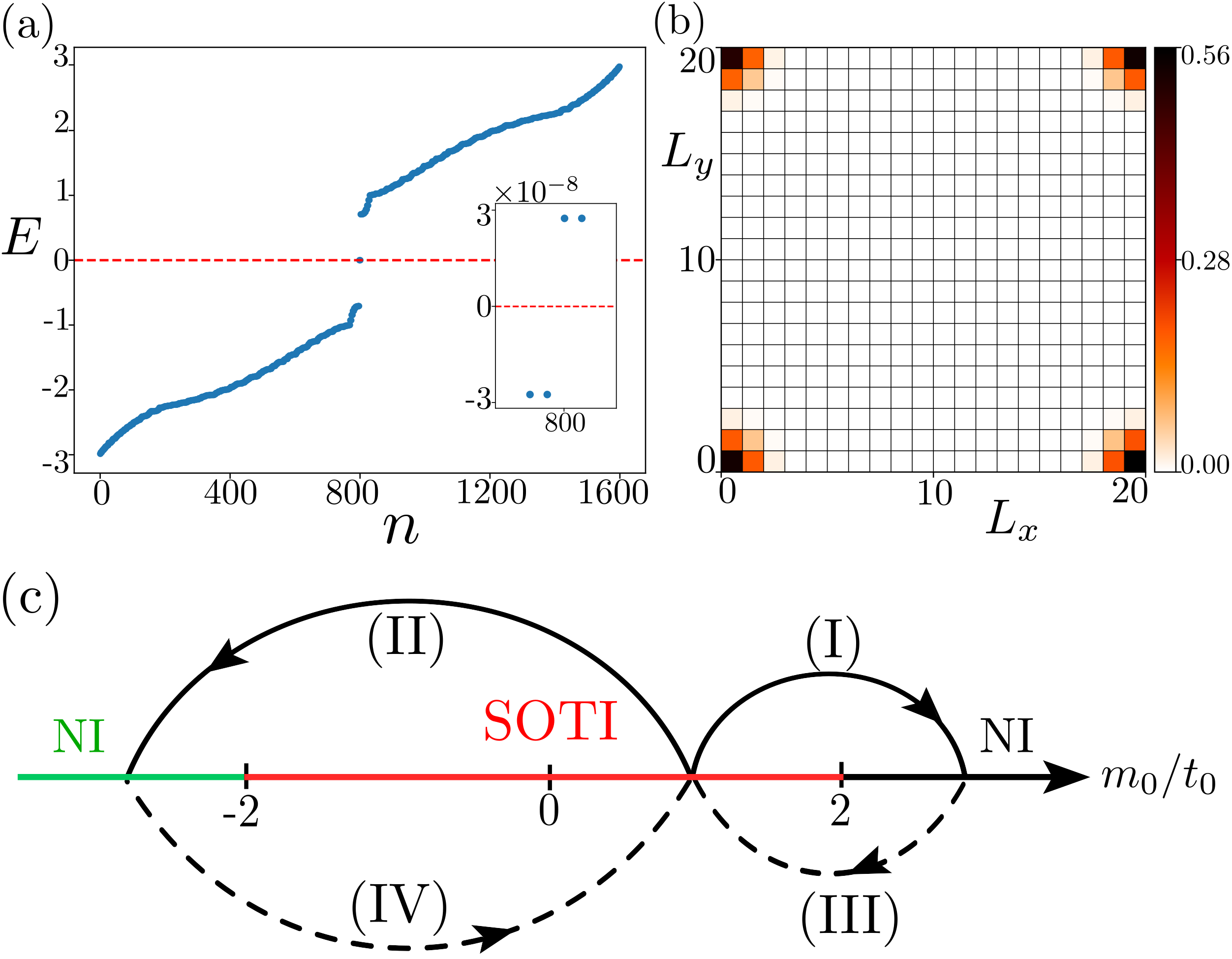}
\caption{(a) Energy ($E$) spectra of the static Hamiltonian [Eq.~\eqref{eq:hamiltonianHOT}] on a square lattice with open boundary conditions in the $x$ and $y$ directions for $t_1=t_0=m_0=\Delta=1$, yielding a second-order topological insulator (SOTI). Here, $n$ is the eigenvalue index. Inset shows four closest to zero-energy modes. (b) The local density of states of those four doubly-degenerate near zero-energy modes is highly localized around four corners of the square lattice. (c) Phase diagram of the Hamiltonian from Eq.~\eqref{eq:hamiltonianHOT}. The real time ramp out of (into) the SOTI phase to (from) normal insulator are shown by solid (dashed) arrows, labeled by Roman numerals. The corresponding dynamic melting (condensation) of the dislocation modes is shown in Figs.~\ref{fig:CornerMeltingProb}-\ref{fig:CornerMeltingLDOS2} (Fig.~\ref{fig:CornerCondensation}). Here, NI stands for normal insulator.   
}~\label{fig:SetupCorner}
\end{figure}

Specifically in $d>1$, it is also conceivable to identify topological crystals harboring gapless modes on boundaries with integer co-dimension $d_c=n>1$, such as the corners and hinges. Together they constitute the family of higher-order topological phases. Namely, an $n$th order topological phase hosts robust topological modes, living on boundaries with $d_c=n$~\cite{HOT:1, HOT:2, HOT:3, HOT:4, HOT:5, HOT:6, HOT:7, HOT:8, HOT:9, HOT:10, HOT:11, HOT:12, HOT:13, HOT:14}. Therefore, in $d=2$ besides the first-order TIs, we also find second-order topological insulators (SOTIs) displaying zero-energy modes localized near four corners of a square lattice, for example, as shown in Fig.~\ref{fig:SetupCorner}.

Therefore, leaving aside their weak cousins~\footnote{The following classification encompasses 2D strong TIs in which the boundary and defect modes are protected by a topological invariant defined on a 2D BZ (in momentum space) or a square lattice (real space), leaving aside the weak TIs that for example can be realized by translational invariant stacking (in the $y$ direction) of $x$ directional one-dimensional topological Su–Schrieffer–Heeger insulators~\cite{SSH:1, SSH:2, SSH:3}. The resulting weak TI (or Dirac semimetal) supports Fermi arc edge modes, but only on the $y$ directional edges. Our results are qualitatively applicable for such topological modes as well.}, the landscape of TIs can be broadly classified into three categories that support edge modes (first-order TIs), dislocation modes (translationally active TIs), and corner modes (SOTIs). Their relationship with the bulk topological invariant is well understood in closed static~\cite{TI:1, TI:2, TI:3, TI:4, TI:5, TI:6, TI:7, TI:8, TI:9, TI:10, TI:11, TI:12, TI:13, TSC:1, TSC:2, TSC:3, TSC:4, TSC:5, TSC:6, defect:1, defect:2, defect:3, defect:4, defect:5, defect:6, defect:7, defect:8, defect:9, defect:10, defect:11, defect:12, defect:13, defect:14, defect:15, HOT:1, HOT:2, HOT:3, HOT:4, HOT:5, HOT:6, HOT:7, HOT:8, HOT:9, HOT:10, HOT:11, HOT:12, HOT:13, HOT:14} and Floquet crystals possessing the time translational symmetry thereby giving birth to the notion of the Floquet BZ~\cite{Floquet:1, Floquet:2, Floquet:3, Floquet:4, Floquet:5, Floquet:6, Floquet:7, Floquet:8, Floquet:9, Floquet:10, Floquet:11, Floquet:12, Floquet:13, Floquet:14}. On the other hand, fate of these modes in open quantum systems that interacts with the environment and described by a time-dependent Hamiltonian devoid of any time translational symmetry is still a subject in its infancy~\cite{OpenTopo:1, OpenTopo:2, OpenTopo:3, OpenTopo:4, OpenTopo:5, OpenTopo:6, OpenTopo:7, OpenTopo:8, OpenTopo:9, OpenTopo:10, OpenTopo:11, OpenTopo:12}, and constitutes the central theme of the present pursuit. Next we summarize our main findings in this context.

\subsection{Summary of main results}

Here we consider square lattice-based 2D open topological quantum systems (see Sec.~\ref{sec:model}) that couple with the environment through the energy channels, mimicking the \emph{weak} measurement protocol. This is an example of quantum non-demolition measurement as we are measuring an observable (energy) that commutes with the Hamiltonian, which makes it a non-disruptive measurement, causing no high energy excitations in the system despite the fact that the system is being monitored by the environment~\cite{OpenTopo:9}. The system-to-environment coupling is parameterized by a positive constant $\gamma$ in the associated Lindblad equation in which the instantaneous time-dependent Hamiltonian plays the role of the Lindblad operator, see Sec.~\ref{subsec:lindbald} and Eq.~\eqref{eq:lindbald}. Such open quantum systems decohere among the energy eigenstates of the instantaneous Hamiltonian, the strength of which is set by $\gamma$. The time-dependent Hamiltonian is modeled by a time-dependent mass parameter for the lattice regularized massive Dirac fermions that smoothly interpolates between topologically distinct insulating phases via continuous quantum phase transitions, see Sec.~\ref{subsec:tdependmass} and Eq.~\eqref{eq:ramprofile}. At the underlying quantum critical point, the band gap closes and the system is described by a collection of massless or gapless Dirac fermions. The ramp speed is parameterized by a positive quantity $\alpha$, which allows us to mimic a slow ramp (for small $\alpha$) as well as a sudden quench (for large $\alpha$). In such systems, we consider two types of time dynamics associated with the topological modes and arrive at the following generic conclusions.

\begin{figure}[t!]
\includegraphics[width=1.00\linewidth]{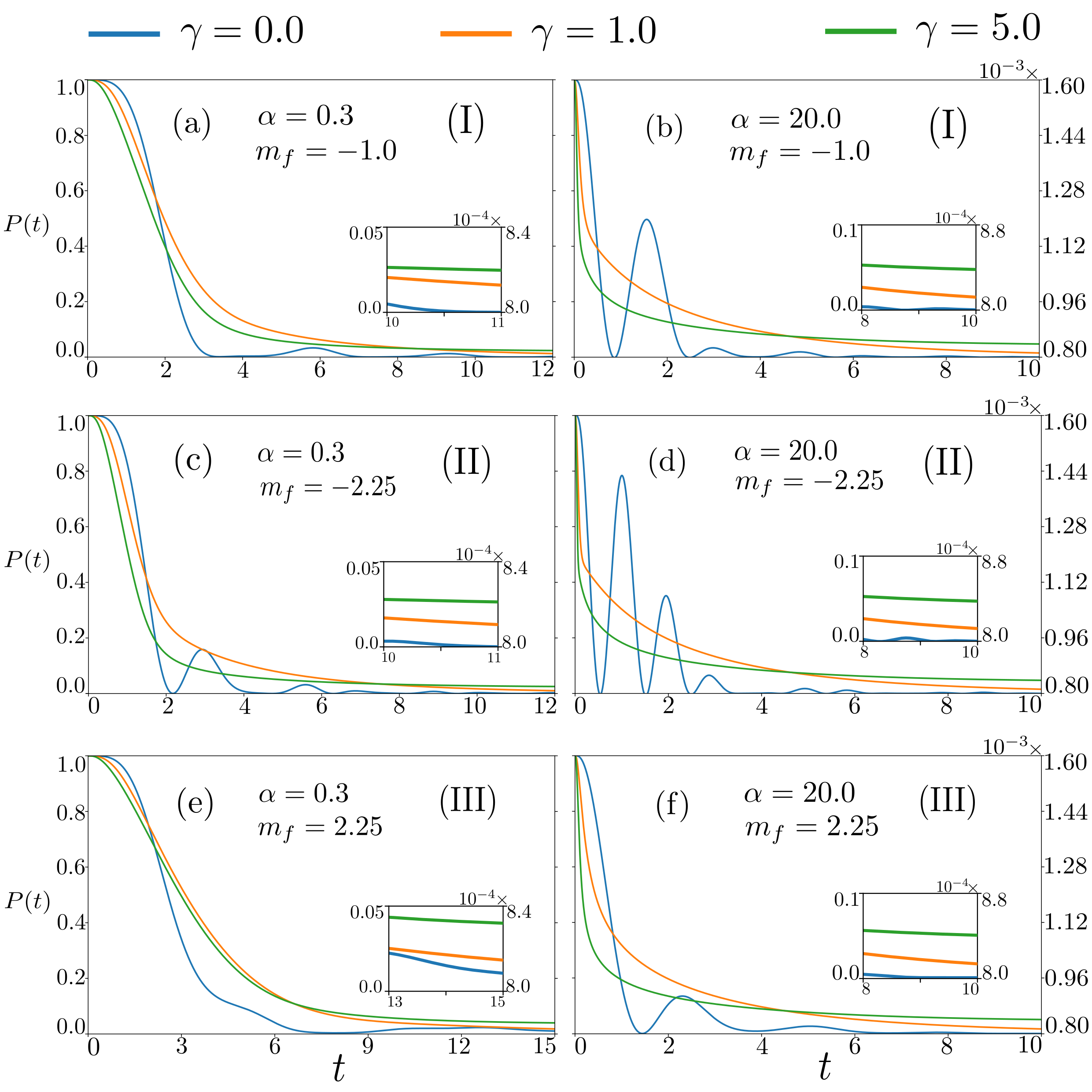}
\caption{Probability $P(t)$ of finding the edge modes [Eq.~\eqref{eq:prob}] in the presence of a real time ramp that takes the system initially in the $\Gamma$ phase at $t=0$ with $m_i=1$ to (a) and (b) the ${\rm M}$ phase with $m_f=-1$, normal insulator living in the proximity to (c) and (d) the ${\rm M}$ phase with $m_f=-2.25$, and (e) and (f) the $\Gamma$ phase with $m_f=2.25$ for a slow ramp with $\alpha=0.3$ [(a), (c), and (e)] and a fast ramp or a sudden quench with $\alpha=20$ [(b), (d), and (f)] for various choices of $\gamma$ quantifying the system-to-environment coupling. See Eqs.~\eqref{eq:ramprofile} and~\eqref{eq:lindbald}. We compute $P(t)$ from a pure state (with the numbers shown on the left vertical axis) and a mixed state ${\rm HF}^\prime$ (with the numbers shown on the right vertical axis). Inset in each subfigure shows the long-time behavior of $P(t)$ for various $\gamma$ values. The real time ramp begins at $t=0$ and thus we show $P(t)$ for $t>0$ only. For $\gamma=0$ we recover the results for the unitary time evolution in a closed system. The Roman numeral in each panel corresponds to the arrow out of the $\Gamma$ phase, see Fig.~\ref{fig:SetupEdge}(c).    
}~\label{fig:EdgeMeltingProb}
\end{figure}

(1) We first consider a situation in which the initial gapped state at time $t=0$ is in the topological regime that features edge or dislocation or corner modes. But the final insulating state realized as $t \to \infty$ is devoid of any such mode.~\footnote{This criterion holds also for edge modes when the real time ramp takes the system out of the $\Gamma$ phase into the ${\rm M}$ phase, see ramp (I) in Fig.~\ref{fig:SetupEdge}(c), as the edge modes in these two phases are distinct. Namely, they propagate in opposite directions and are protected by distinct first Chern numbers.} Then, soon after the ramp is switched on, the signature of all the topological modes starts to decrease more rapidly with increasing system-to-environment coupling ($\gamma$). This outcome can be anticipated from the fact that with larger value of $\gamma$ the initial topological modes encounter stronger decoherence among all the other instantaneous energy eigenbasis of the instantaneous Hamiltonian (including the bulk states), constituting the decoherence channels made available by weak measurements. However, with increasing time as the system approaches the quantum critical point across which the initial topological mode ceases to exist, it freezes due to the critical slow down caused by the diverging of associated time scales and falls out of equilibrium. Then the quantum Zeno effect sets in, stabilizing the state and slowing down the eventual decay. This mechanism becomes more prominent with increasing $\gamma$. Consequently, at long time scale the probability of finding the initial topological mode (although small) is larger for larger $\gamma$. Nevertheless, such a probability ultimately vanishes as the final state does not host the initial topological mode or is completely devoid of it. Although these results are qualitatively insensitive to the ramp speed ($\alpha$), the quantum Zeno effect sets in at shorter time scale with increasing $\alpha$, as then the system arrives at the topological quantum critical point, harboring nodal Dirac fermions, more rapidly. We arrive at these conclusions from the time evolution of the probability of finding these topological modes and their site-resolved local density of states (LDOS) at various time instants. The results are shown in Figs.~\ref{fig:EdgeMeltingProb}-\ref{fig:CornerMeltingLDOS2}. We refer to this phenomenon as the \emph{dynamic melting} of topological modes, discussed in Sec.~\ref{subsec:Tdynamicsresultsmelting}.

(2) In the next step, we consider a reverse course of the time evolution in which the initial gapped system at $t=0$ is devoid of any topological mode but the real time ramp takes it into an insulating phase as $t \to \infty$ that harbors topologically robust edge or dislocation or corner modes at least in the static limit.~\footnote{This criterion is operative for the edge modes in the final $\Gamma$ phase when the real time ramp begins in the ${\rm M}$ phase which also supports edge modes, see ramp (IV) in Fig.~\ref{fig:SetupEdge}(c). This is so because the edge modes in the $\Gamma$ and ${\rm M}$ phases are protected by distinct first Chern number and travel in the opposite directions.} Then, with increasing $\gamma$ the probability of finding the topological modes increases in the final state, irrespective of the ramp speed ($\alpha$). This generic observation stems from the fact that the initial state decoheres among all the eigenstates of the instantaneous Hamiltonian strongly with increasing $\gamma$, which also leads to the enrichment of the topological modes in the final state once the system crosses the underlying quantum critical point. Such a strong decoherence gives birth to an enhanced probability of finding the topological modes in the final insulating phase. Nevertheless, with a slower ramp (equivalently smaller $\alpha$) such a probability gets amplified, possibly following the adiabatic theorem. The results are shown in Figs.~\ref{fig:EdgeCondensation}-\ref{fig:CornerCondensation}, depicting the time evolution of the probability of finding these modes and their site-resolved LDOS at various instants of time. We refer to this phenomenon as the \emph{dynamic condensation} of topological modes, discussed in Sec.~\ref{subsec:Tdynamicsresultscondensation}.

\subsection{Organization}

The rest of the paper is organized as follows. In the next section (Sec.~\ref{sec:model}) we introduce the square lattice-based model Hamiltonian for various types of TIs (first-order, translationally active, and second-order), realizable in two spatial dimensions. In Sec.~\ref{sec:Tdynamicsformalism}, we develop and discuss all the requisite theoretical tools to compute the time dynamics of topological modes in open quantum crystals. Section~\ref{sec:Tdynamicsresults} showcases all the outcomes related to the dynamic melting and condensation of the edge, dislocation, and corner modes. Our findings are summarized in Sec.~\ref{sec:summary}, where we also discuss possible experimental platforms to test our predictions and outline some future generalizations of the current work.   

\section{Planar TI$\text{s}$: Model Hamiltonian}~\label{sec:model}

2D quantum crystals can foster various incarnations of TIs, among which strong and translationally active ones, belonging to the family of first-order TIs, and SOTIs, belonging to the family of higher-order TIs are the most prominent ones. We begin the discussion by introducing their model Hamiltonian on a square lattice in three subsequent subsections, respectively.

\begin{figure}[t!]
\includegraphics[width=1.00\linewidth]{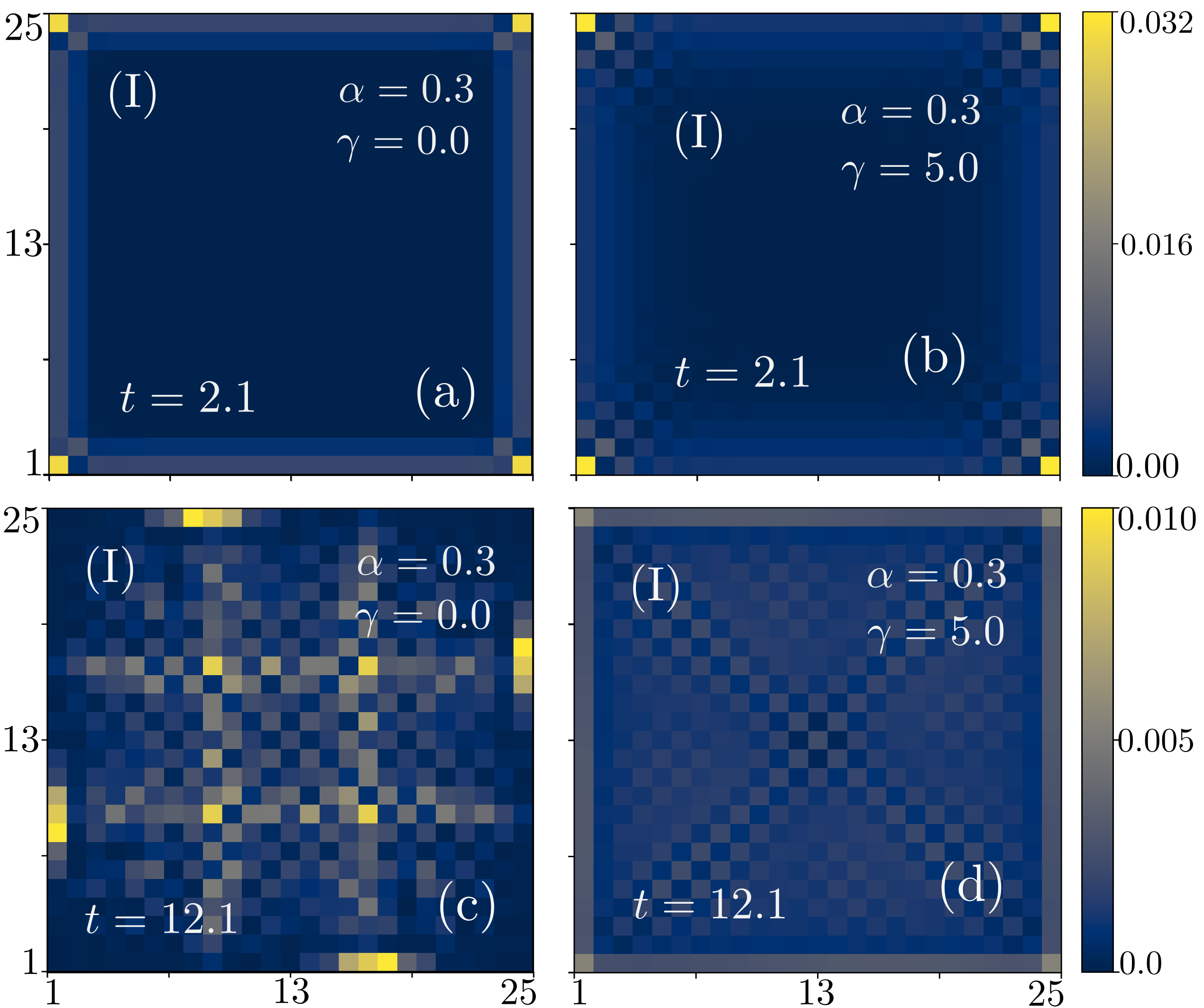}
\caption{Site-resolved local density of states (LDOS), computed from the density matrix $\rho(t)$ [see Eq.~\eqref{eq:LDOS}] when the initial density matrix $\rho(0)$ is constructed from a single edge mode (pure state) for a slow ramp ($\alpha=0.3$) at (a) and (b) an earlier time instant $t=2.1$ and (c) and (d) at a later time instant $t=12.1$. The left panels [(a) and (c)] correspond to a unitary time evolution in a closed system with $\gamma=0$, while the right panels [(b) and (d)] correspond to a non-unitary time evolution in an open system with the system-to-environment coupling strength $\gamma=5.0$. See Eqs.~\eqref{eq:ramprofile} and~\eqref{eq:lindbald}. Here we choose $m_i=1.0$ and $m_f=-1.0$. We arrive at the same profile for the site-resolved LDOS when computed from a mixed state ${\rm HF}^\prime$ once the uniform background LDOS for the half-filled system is subtracted. The corresponding time evolution of the probability of finding the edges mode $P(t)$ is shown in Fig.~\ref{fig:EdgeMeltingProb}(a). The Roman numeral in each panel corresponds to the arrow out of the $\Gamma$ phase, shown in Fig.~\ref{fig:SetupEdge}(c). In panel (c), the LDOS shows some peaks near the edges of the system as the final state (${\rm M}$ phase) also fosters edge modes, which are, however, distinct from the ones in the initial state ($\Gamma$ phase).   
}~\label{fig:EdgeMeltingLDOS1}
\end{figure}

\subsection{First-order TI}

We first consider the model Hamiltonian for 2D first-order TIs on a square lattice. Here we focus on one of their simplest representatives, the QAHI. The associated massive Dirac Hamiltonian, namely the lattice-regularized Qi-Wu-Zhang model~\cite{QWZ}, takes the form $H_1=\tau_x d_x(\vec{k}) + \tau_y d_y(\vec{k}) + \tau_z d_z(\vec{k})$, explicitly given by
\allowdisplaybreaks[4]
\begin{equation}~\label{eq:hamiltonian}
H_1=t_1 \sum_{j=x,y} \sin(k_j a) \tau_j + \bigg( t_0 \sum_{j=x,y} \cos(k_j a) -m_0 \bigg) \tau_z.
\end{equation}       
The vector Pauli matrix ${\boldsymbol \tau}=(\tau_x,\tau_y,\tau_z)$ operates on the orbital indices and $\vec{k}=(k_x, k_y)$ is the momentum. This model supports topological and trivial insulating phases for $|m_0/t_0|<2$ and $|m_0/t_0|>2$, respectively. In the entire topological regime this model features zero-energy modes that are localized near the edges of the square lattice. It can be verified by numerically diagonalizing the real space Hamiltonian associated with $H_1$ on a square lattice with open boundary conditions in both directions, obtained from Eq.~\eqref{eq:hamiltonian} via Fourier transformations, and subsequently computing the LDOS of two closest to zero-energy modes. The results are shown in Fig.~\ref{fig:SetupEdge}. In this case a two-dimensional ($d=2$) TI supports robust gapless modes on one-dimensional ($d_b=1$) boundaries or edges. Hence, QAHI is a first-order ($n=1=d_c$) TI, supporting gapless modes on boundaries with co-dimension $d_c=d-d_b=2-1=1$.

\begin{figure}[t!]
\includegraphics[width=1.00\linewidth]{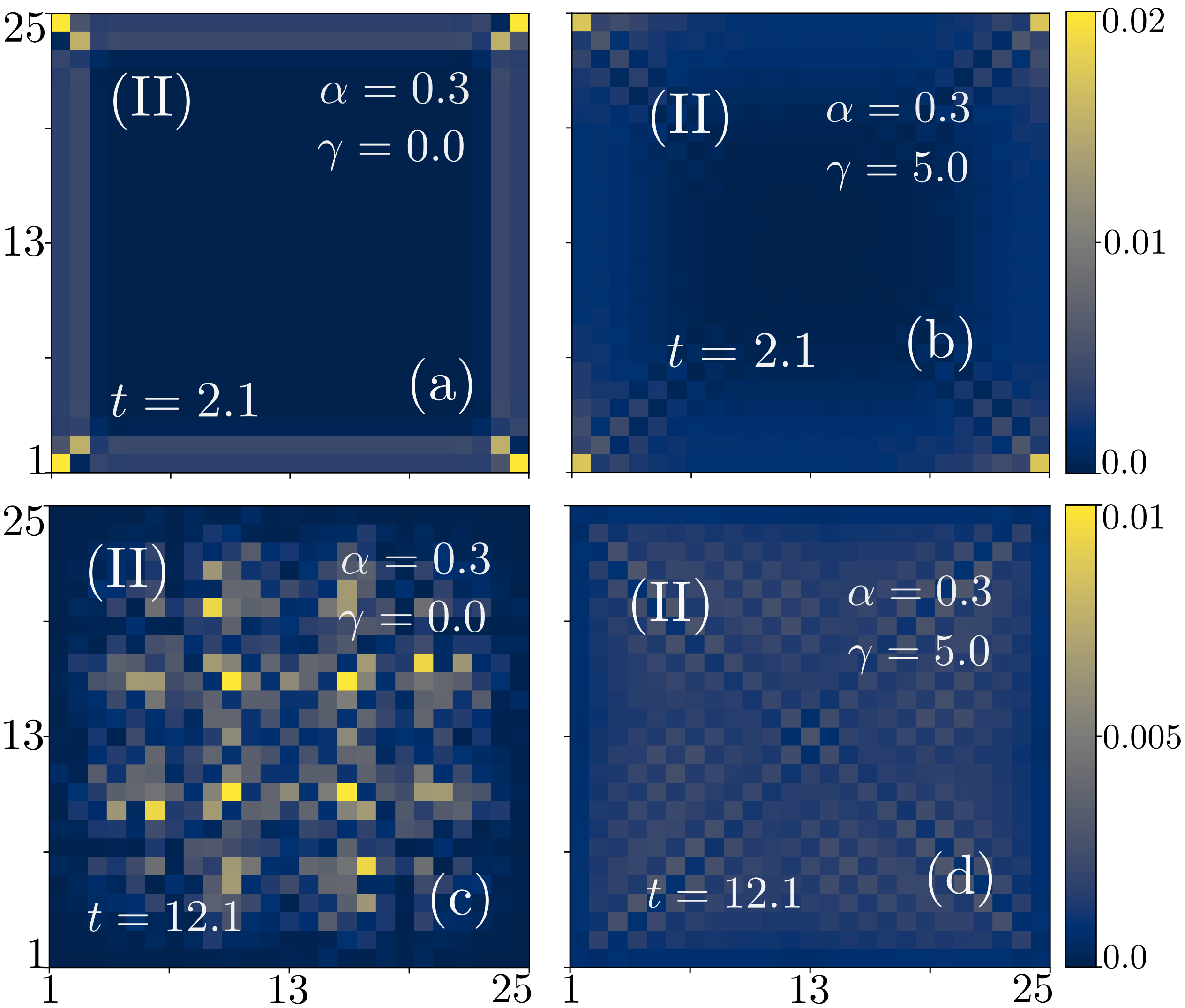}
\caption{Same as Fig.~\ref{fig:EdgeMeltingLDOS1}, but for $m_i=1.0$ and $m_f=-2.25$. The corresponding time evolution of the probability of finding the edge modes at any time $t$ is shown in Fig.~\ref{fig:EdgeMeltingProb}(c). The Roman numeral in each panel corresponds to the arrow out of the $\Gamma$ phase, shown in Fig.~\ref{fig:SetupEdge}(c).     
}~\label{fig:EdgeMeltingLDOS2}
\end{figure}

These edge modes are topological in nature, ensured by the nontrivial bulk topological invariant, the first Chern number ($C$) for the filled valence band, defined as  
\allowdisplaybreaks[4] 
\begin{equation}~\label{eq:chernnumber}
C=-\int_{\rm BZ} \dfrac{d^{2}{\vec k}}{4\pi} \:\: \big[ \partial_{k_x} \hat{\vec{d}}(\vec{k}) \times \partial_{k_y} \hat{\vec{d}}(\vec{k}) \big] \cdot \hat{\vec{d}}(\vec{k}),
\end{equation}
where $\hat{\vec{d}}(\vec{k})=\vec{d}(\vec{k})/|\vec{d}(\vec{k})|$~\cite{TKNN}. The integral is restricted within the first BZ. However, the topological regime fragments into two sectors depending on the TRIM point in the BZ around which the band inversion takes place. As such for $0<m_0/t_0<2$ the band inversion takes place around the $\Gamma$ point (the $\Gamma$ phase), whereas for $-2<m_0/t_0<0$ the band inversion takes place around the ${\rm M}$ point (the ${\rm M}$ phase) of the BZ. Both the $\Gamma$ and ${\rm M}$ phases represent strong TIs with $C=+1$ and $-1$, respectively, and the corresponding edge modes encode the bulk-boundary correspondence.

Note that there exists an antiunitary operator $\Theta=\tau_x {\mathcal K}$, where ${\mathcal K}$ is the complex conjugation operator, such that $\{H_1,\Theta \}=0$. Then two closest to zero-energy edge modes, denoted by $\ket{\Psi^{\rm edge}_{1,2}}$, living on either side of zero energy, are related to each other according to $\ket{\Psi^{\rm edge}_{1,2}}=\Theta \ket{\Psi^{\rm edge}_{2,1}}$. As such the eigenvectors of any pair of states of $H_1$ at energies $\pm E$ are related by $\ket{\mp E}=\Theta \ket{\pm E}$. These relationships hold up to an overall unimportant phase. Therefore, $\Theta$ is the generator of the anti-unitary particle-hole symmetry of $H_1$.

\subsection{Translationally active TI}

The Qi-Wu-Zhang Hamiltonian [Eq.~\eqref{eq:hamiltonian}] also allows us to identify a translationally active TI within the territory of first-order TIs. Notice, although both $\Gamma$ and ${\rm M}$ phases support gapless edge modes, they are topologically distinct phases of matter, characterized by distinct first Chern numbers $C$ [Eq.~\eqref{eq:chernnumber}]. Specifically, the finite band inversion momentum in the ${\rm M}$ phase can be probed in terms of topologically robust dislocation modes. A dislocation lattice defect can be constructed via the Volterra cut and paste procedure from a square lattice, which has already been discussed in books and literature~\cite{defect:0, defect:1, defect:2, defect:3, defect:4, defect:5, defect:6, defect:7, defect:8, defect:9, defect:10, defect:11, defect:12, defect:13, defect:14, defect:15}. So, we do not repeat it here.

A dislocation lattice defect is characterized by the underlying Burgers vector ${\bf b}$, measuring the missing translation near the core of such a defect. Consequently, when an electron with momentum ${\bf K}$ encircles a dislocation core it picks up an extra hopping phase $\exp[i \Phi]$, where $\Phi={\bf K} \cdot {\bf b}$. In a topological phase with the band inversion at momentum ${\bf K}_{\rm inv}$, this phase is given by $\Phi={\bf K}_{\rm inv} \cdot {\bf b}$, and thus $\Phi=0$ and $\pi$ in the $\Gamma$ and ${\rm M}$ phases, respectively. When massive Dirac fermions experience a $\pi$ flux, as is the case around the dislocation core in the ${\rm M}$ phase, it underpins localized zero-energy modes around the defect core, manifesting the bulk-defect correspondence. We verify this prediction by numerically diagonalizing the real space Hamiltonian with a pair of dislocation-antidislocation in an otherwise square lattice with periodic boundary conditions in both directions, see Fig.~\ref{fig:SetupDislocation}.

\begin{figure}[t!]
\includegraphics[width=1.00\linewidth]{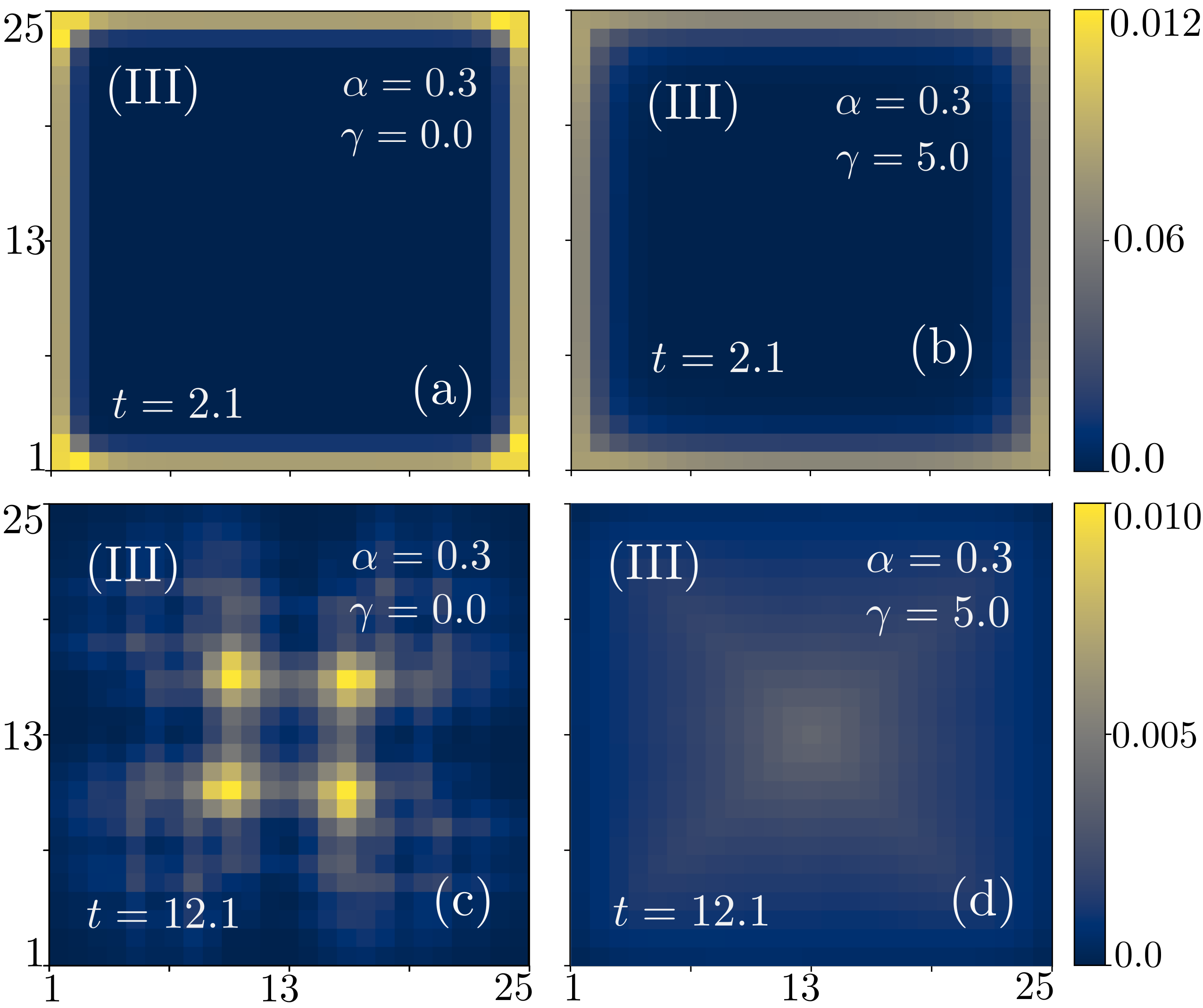}
\caption{Same as Fig.~\ref{fig:EdgeMeltingLDOS1}, but for $m_i=1.0$ and $m_f=2.25$. The corresponding time evolution of the probability of finding the edge modes at any time $t$ is shown in Fig.~\ref{fig:EdgeMeltingProb}(e). The Roman numeral in each panel corresponds to the arrow out of the $\Gamma$ phase, shown in Fig.~\ref{fig:SetupEdge}(c).      
}~\label{fig:EdgeMeltingLDOS3}
\end{figure}

As the existence of the ${\rm M}$ point in the BZ rests on the translational symmetry of the underlying square lattice, the ${\rm M}$ phase is also coined as \emph{translationally active}, which can be identified from the gapless modes bound to dislocation lattice defects, constructed by locally breaking the translational symmetry of the parent square lattice. However, such gapless dislocation modes do not exist in the $\Gamma$ phase, as $\Phi=0$ therein, which is thus translationally inert. The anti-unitary particle-hole symmetry, generated by $\Theta=\tau_x {\mathcal K}$, also relates two closest to zero-energy dislocation modes, denoted by $\ket{\Psi^{\rm dis}_{1,2}}$, living on either side of the zero-energy, according to $\ket{\Psi^{\rm dis}_{1,2}}=\Theta \ket{\Psi^{\rm dis}_{2,1}}$, up to an overall unimportant phase.

\begin{figure}[t!]
\includegraphics[width=1.00\linewidth]{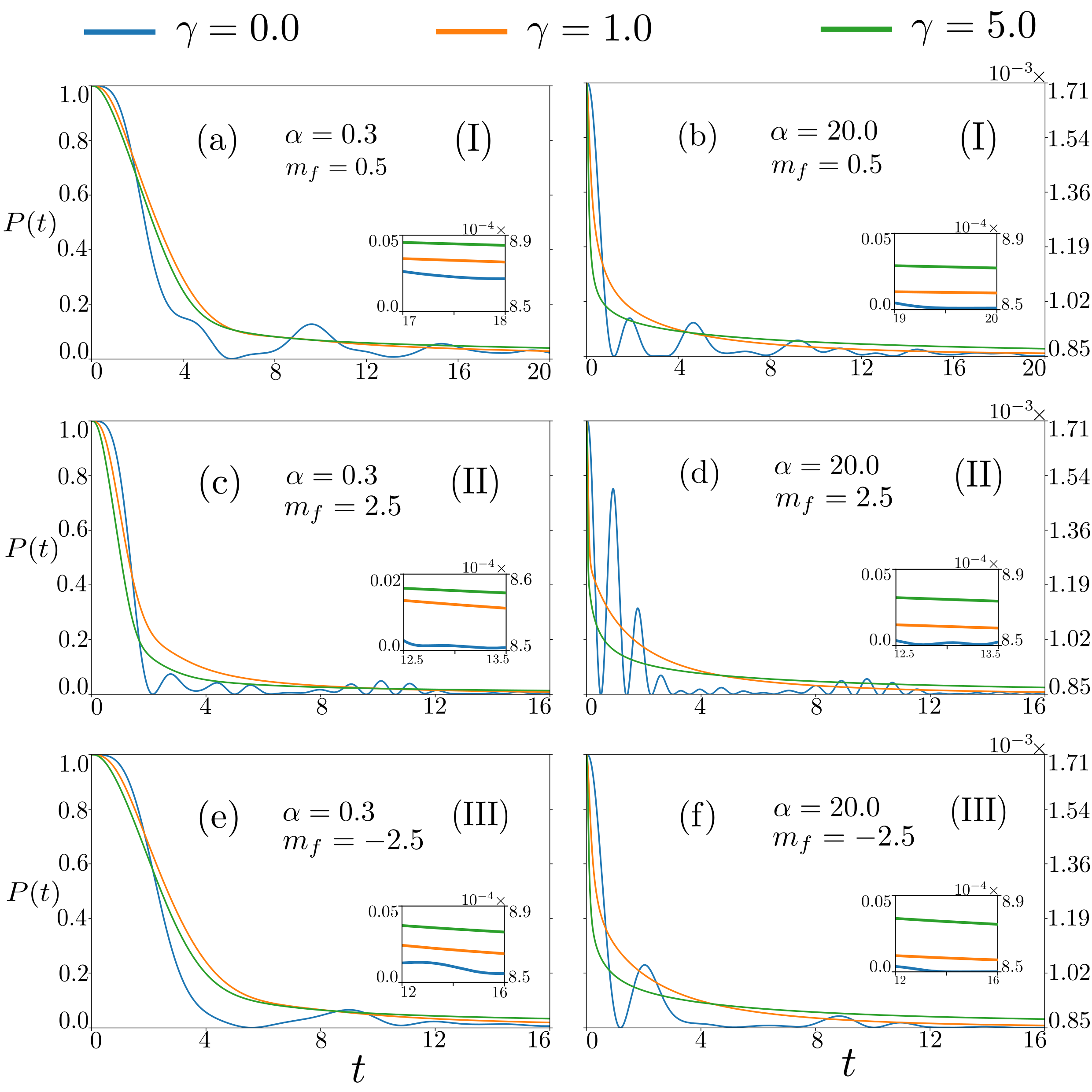}
\caption{Probability $P(t)$ of finding the dislocation modes [Eq.~\eqref{eq:prob}] in the presence of a real time ramp that takes the system initially in the translationally active ${\rm M}$ phase at $t=0$ with $m_i=-1.0$ to (a) and (b) the $\Gamma$ phase with $m_f=0.5$, normal insulator living in the proximity to (c) and (d) the $\Gamma$ phase with $m_f=2.25$, and (e) and (f) the ${\rm M}$ phase with $m_f=-2.25$ for a slow ramp with $\alpha=0.3$ [(a), (c), and (e)] and a fast ramp or a sudden quench with $\alpha=20$ [(b), (d), and (f)] for various choices of $\gamma$ quantifying the system-to-environment coupling. See Eqs.~\eqref{eq:ramprofile} and~\eqref{eq:lindbald}. We compute such probability from a pure state (with the numbers shown on the left vertical axis) and a mixed state ${\rm HF}^\prime$ (with the numbers shown on the right vertical axis). Inset in each subfigure shows the long-time behavior of $P(t)$ for various $\gamma$ values. The real time ramp begins at $t=0$ and thus we show $P(t)$ for $t>0$ only. For $\gamma=0$ we recover the unitary time evolution in a closed system. The Roman numeral in each panel corresponds to the arrow out of the ${\rm M}$ phase, shown in Fig.~\ref{fig:SetupDislocation}(c).   
}~\label{fig:DislocationMeltingProb}
\end{figure}

\subsection{Higher-order TIs}

In two spatial dimensions, it is also conceivable to realize a SOTI that fosters topologically robust zero-energy modes at four corner of a square lattice, characterized by the co-dimension $d_c=2-0=2$. The Bloch Hamiltonian describing such a phase of matter is given by 
\allowdisplaybreaks[4]
\begin{eqnarray}~\label{eq:hamiltonianHOT}
H_2 &=& t_1 \sum_{j=x,y} \sin(k_j a) \Gamma_j + \bigg( t_0 \sum_{j=x,y} \cos(k_j a) -m_0 \bigg) \Gamma_z \nonumber \\
&+& \Delta \left[ \cos(k_x a) -\cos(k_y a) \right] \Gamma_{x^2-y^2},
\end{eqnarray}       
where $\Gamma_x=\sigma_z \otimes \tau_x$, $\Gamma_y=\sigma_z \otimes \tau_y$, $\Gamma_z=\sigma_z \otimes \tau_z$, and $\Gamma_{x^2-y^2}=\sigma_x \otimes \tau_0$ are four mutually anti-commuting four-dimensional Hermitian matrices, each of which squares to unity, and $\otimes$ corresponds to the Kronecker product. The newly introduced Pauli matrices $\{ \sigma_\mu \}$ operate on the spin indices. Here, $\tau_0$ and $\sigma_0$ are two-dimensional identity matrices. For $\Delta=0$, the above Hamiltonian describes the Bernevig-Hughes-Zhang model for quantum spin Hall insulator~\cite{TI:4}. It supports two counter-propagating helical edge modes for opposite spin projections when $|m_0/t_0|<2$. A finite $\Delta$ causes hybridization between them, and gaps out these edge modes, but only partially. Specifically, the edge modes get gapped everywhere except at four corners of the square lattice, residing on its diagonals $\pm k_x=\pm k_y$ along which $\cos(k_x a) -\cos(k_y a)=0$. Therefore, within the topological regime of the Bernevig-Hughes-Zhang model, any finite $\Delta$ gives birth to a SOTI, accommodating four near zero-energy modes that are highly localized around four corners of a square lattice with open boundary conditions, as shown in Fig.~\ref{fig:SetupCorner}. It should be noted that with some redefinition of the parameters and four mutually anticommuting four-dimensional Hermitian matrices, appearing in Eq.~\eqref{eq:hamiltonianHOT}, $H_2$ maps onto the original Benalcazar-Bernevig-Hughes model for SOTI in two dimensions~\cite{HOT:1, HOT:12}.

The Hamiltonian $H_2$ for $\Delta=0$, preserves the time-reversal symmetry ($\mathcal T$), generated by ${\mathcal T}=\sigma_y \otimes \tau_x {\mathcal K}$ with ${\mathcal T}^2=-1$. It ensures the two-fold Kramers' degeneracy of all the states. The Bernevig-Hughes-Zhang model also preserves the parity (${\mathcal P}$) symmetry, generated by $\Gamma_z$, under which the momentum $\vec{k} \to -\vec{k}$. On the other hand, when $\Delta$ is finite, the Hamiltonian $H_2$ breaks both ${\mathcal T}$ and ${\mathcal P}$ symmetries, but preserves the composite ${\mathcal P}{\mathcal T}$ symmetry, which now guarantees the two-fold degeneracy of all the eigenstates. For any finite $\Delta$, the Bloch Hamiltonian $H_2$ breaks the four-fold ($C_4$) rotational symmetry about the $z$-direction, generated by $\exp\left[i \pi \Gamma_{xy}/4 \right]$ under which $(k_x,k_y) \to (-k_y,k_x)$, where $\Gamma_{xy}=i \Gamma_x \Gamma_y$. Therefore, the term proportional to $\Delta$ corresponds to a discrete symmetry breaking Wilson-Dirac mass and each one of the four zero-energy corner modes breaks the $C_4$ symmetry. It is worth noting that $H_2$ preserves two additional composite symmetries, namely $C_4{\mathcal T}$ and $C_4{\mathcal P}$~\cite{defect:8}.

Notice that $\{ H_2, \Gamma_5 \}=0$, where $\Gamma_5=\sigma_y \otimes \tau_0$ is the generator of the unitary particle-hole or sublattice or chiral symmetry. If we denote four near zero-energy corner modes by $\ket{\Psi^{\rm cor}_j}$ with $j=1, \cdots, 4$, among which two degenerate states at sufficiently small positive (negative) energy are $\ket{\Psi^{\rm cor}_j}$ with $j=1,2$ ($\ket{\Psi^{\rm cor}_j}$ with $j=3,4$), then $\ket{\Psi^{\rm cor}_{1,2}}=\Gamma_5 \ket{\Psi^{\rm cor}_{3,4}}$ and vice-versa. As such the eigenvectors of any pair of states of $H_2$ at energies $\pm E$ are related by $\ket{\mp E}=\Gamma_5 \ket{\pm E}$. These relationships hold up to an overall unimportant phase.

As the penultimate remark of this section, we point out that it is because of these particle-hole symmetries (anti-unitary or unitary) of $H_1$ and $H_2$, the electronic density in a half-filled system displays a uniform distribution throughout the system irrespective of the presence or absence of any topological modes~\cite{OpenTopo:10}. This observation will turn out to be crucial when we discuss the time dynamics of the topological modes in terms of the density matrices, about which more in the next section.

\begin{figure}[t!]
\includegraphics[width=1.00\linewidth]{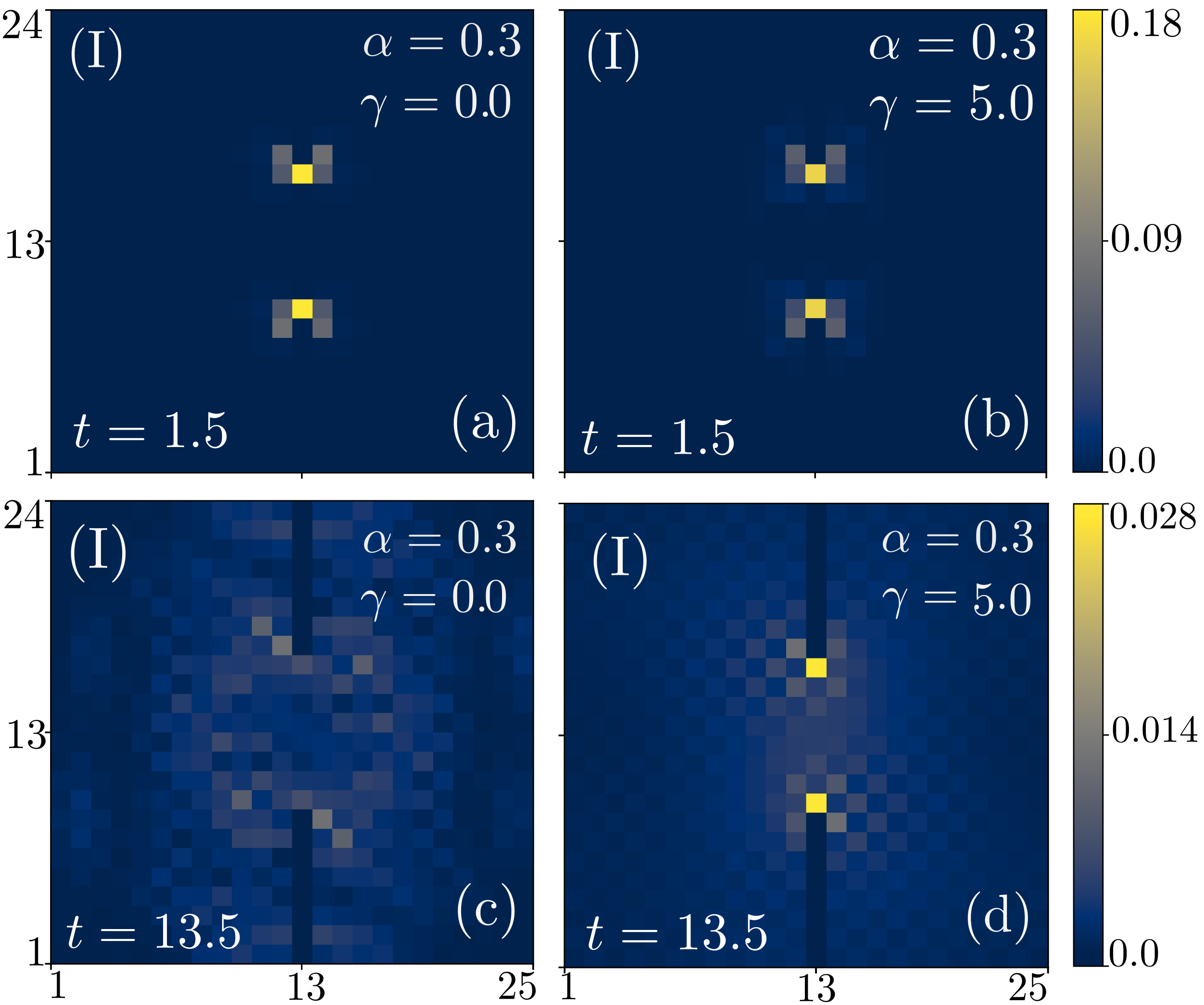}
\caption{Site-resolved local density of states (LDOS), computed from the density matrix $\rho(t)$ [see Eq.~\eqref{eq:LDOS}] when the initial density matrix $\rho(0)$ is constructed from a single dislocation mode (pure state) for a slow ramp with $\alpha=0.3$ at (a) and (b) an earlier time instant $t=1.5$, and (c) and (d) a later time instant $t=13.5$. The left panels [(a) and (c)] correspond to a unitary time evolution in a closed system with $\gamma=0$, while right panels [(b) and (d)] correspond to a non-unitary time evolution in an open system with the system-to-environment coupling strength $\gamma=5.0$. See Eqs.~\eqref{eq:ramprofile} and~\eqref{eq:lindbald}. Here we choose $m_i=-1.0$ and $m_f=0.5$. We arrive at the same profile for the site-resolved LDOS when computed from a mixed state ${\rm HF}^\prime$ once the uniform background LDOS for the half-filled system is subtracted. The corresponding time evolution of the probability of finding the dislocation modes at any time $t$ is shown in Fig.~\ref{fig:DislocationMeltingProb}(a). The Roman numeral in each panel corresponds to the arrow out of the ${\rm M}$ phase, shown in Fig.~\ref{fig:SetupDislocation}(c).  
}~\label{fig:DislocationMeltingLDOS1}
\end{figure}

Finally, we note that any two topologically distinct insulators are separated by a quantum critical point. The continuous quantum phase transition between them takes place through a band gap closing at a high symmetry point in the BZ.  On the $m_0/t_0$ axis the topological quantum critical points are located at $m_0/t_0=2,0$, and $-2$ for the lattice-regularized Qi-Wu-Zhang model. At these points, the band gap closes at the $\Gamma$, ${\rm X}=(\pi,0)/a$ and ${\rm Y}=(0,\pi)/a$, and ${\rm M}$ points, respectively. By contrast, the band gap in the spectrum of $H_2$ [see Eq.~\eqref{eq:hamiltonianHOT}], harboring SOTI besides the normal ones, only closes at $m_0/t_0=\pm 2$. The universality class of all such topological phase transitions is captured by a collection of massless Dirac fermions for which the Hamiltonian is given by 
\begin{equation}
H_{\rm QCP} = t_1 \left( \sin(k_x a) N_x + \sin(k_y a) N_y \right).
\end{equation}
The Hermitian matrices $N_x= \sigma_x$ and $N_y=\sigma_y$ near the first-order (strong and translationally active) to trivial insulator quantum phase transition. On the other hand, $N_x= \Gamma_x$ and $N_y=\Gamma_y$ at the quantum critical point separating a SOTI from a normal or trivial insulator. Notice that the SOTIs realized for $-2<m_0/t_0<0$ and $0<m_0/t_0<2$, although not separated by any band gap closing at $m_0/t_0 =0$, are distinct phases of matter as only the former one supports robust dislocation modes and thus also translationally active. However, such defect modes are typically placed at finite energies, which are protected by the composite ${\mathcal P} {\mathcal T}$, $C_4{\mathcal T}$, and $C_4{\mathcal P}$ symmetries~\cite{defect:8, defect:12}. In this work, we do not delve into the dynamics of dislocation modes in a SOTI. Irrespective of these details, the universality class of all the topological quantum phase transitions is set by the dynamic scaling exponent $z=1$, measuring the relative scaling between energy and momentum according to $E \sim |\vec{k}|^z$ in gapless Dirac systems, and the correlation length exponent $\nu=1$, resulting from the scaling dimension of the mass parameter $m_0$ therein~\cite{TQPT:1, TQPT:2}.

\begin{figure}[t!]
\includegraphics[width=1.00\linewidth]{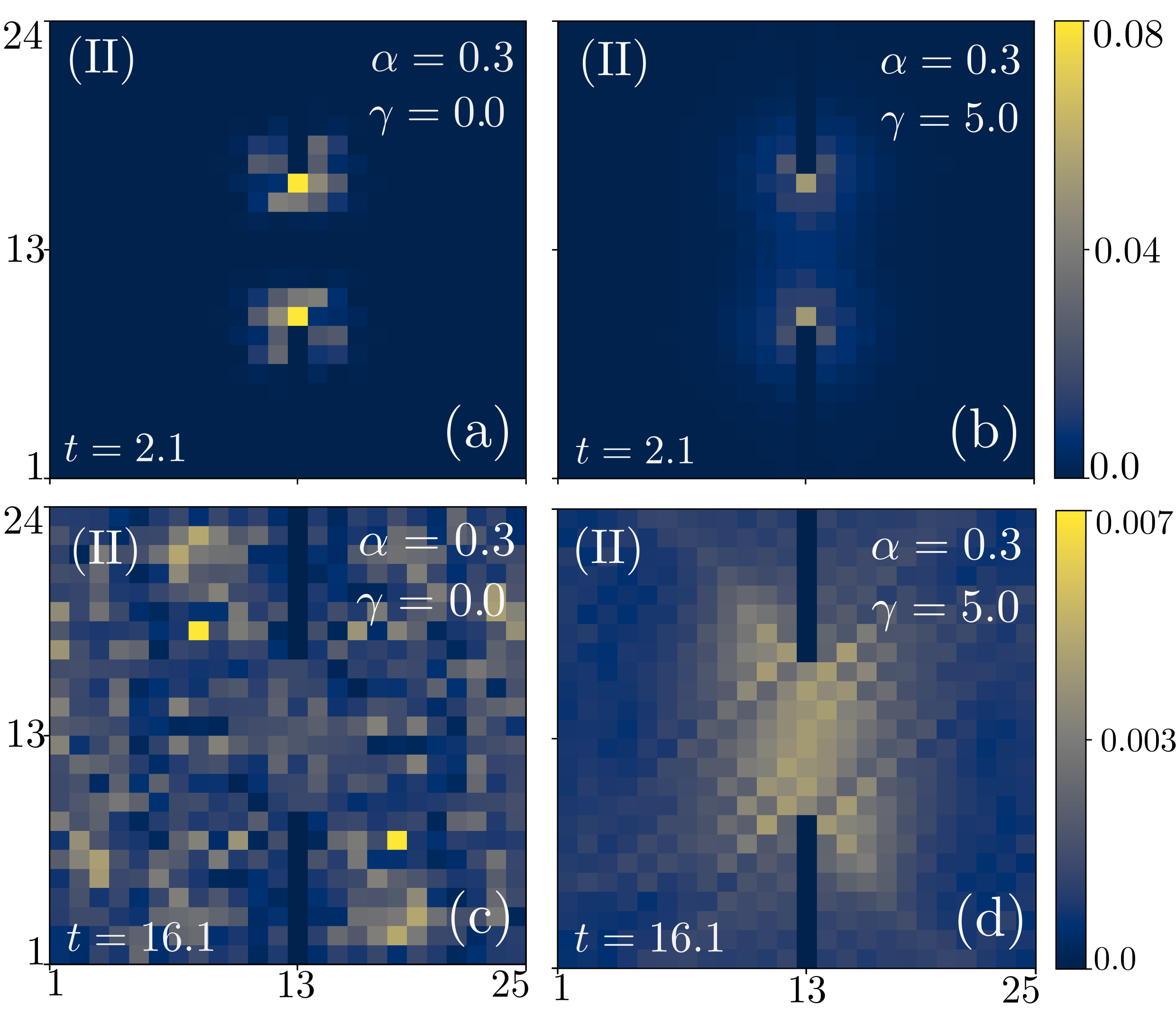}
\caption{Same as Fig.~\ref{fig:DislocationMeltingLDOS1}, but for $m_i=-1.0$ and $m_f=2.5$. The corresponding time evolution of the probability of finding the dislocation modes at any time $t$ is shown in Fig.~\ref{fig:DislocationMeltingProb}(c). The Roman numeral in each panel corresponds to the arrow out of the ${\rm M}$ phase, shown in Fig.~\ref{fig:SetupDislocation}(c).     
}~\label{fig:DislocationMeltingLDOS2}
\end{figure}

\section{Time dynamics: Formalism}~\label{sec:Tdynamicsformalism}

In this section, we develop the requisite theoretical framework to scrutinize the time dynamics of the edge, dislocation, and corner modes, discussed in the previous section in open quantum systems. First, we promote the real time ramp protocol that gives rise to a time-dependent Hamiltonian smoothly interpolating between any pair of topologically distinct insulators (Sec.~\ref{subsec:tdependmass}). Subsequently, we discuss the formalism to describe such an open quantum system that decoheres among various energy eigenstates of the instantaneous Hamiltonian in terms of the Lindblad equation (Sec.~\ref{subsec:lindbald}). Finally, we outline the density matrices we consider to investigate the quantum dynamics of various topological modes in open quantum systems (Sec.~\ref{subsec:densitymatrix}).

\subsection{Time-dependent Hamiltonian}~\label{subsec:tdependmass}

A real time ramp is modeled by taking the mass parameter $m_0 \to m(t)$ in Eqs.~\eqref{eq:hamiltonian} and~\eqref{eq:hamiltonianHOT}, where  
\allowdisplaybreaks[4]
\begin{equation}~\label{eq:ramprofile}
m(t)= m_i+ \left( m_f-m_i \right) \: \left[ 1- \exp(-\alpha t) \right],
\end{equation} 
and the resulting time-dependent Hamiltonian is denoted by $H_1(t)$ and $H_2(t)$, respectively. The real positive parameter $\alpha$ sets the speed of the ramp. Namely for a small (sufficiently large) $\alpha$ we realize a slow ramp (sudden quench). Note that $m(0)=m_i$ sets the initial value of the mass parameter at the beginning of the ramp at $t=0$ and $m(t \to \infty)=m_f$ yields the final value of the mass parameter at the end of the real time ramp. It allows us to smoothly interpolate between various insulating phases, discussed in Sec.~\ref{sec:model}, and we characterize each ramp by a set of parameters $m_i$, $m_f$, and $\alpha$~\cite{OpenTopo:2}. Although not directly pertinent, the rate of change of the mass parameter is given by
\begin{equation}
\frac{d m(t)}{d t} = \alpha \; (m_f-m_i) \; \exp[-\alpha t].
\end{equation}
Therefore, the time variation of the mass term is large initially close to $t=0$, but becomes progressively smaller with increasing $t$. This quantity is also larger for higher ramp speed ($\alpha$) and for the increasing distance between the initial and final insulating states in the parameter space, characterized by $m_i$ and $m_f$, respectively. Strictly, a time-dependent Hamiltonian can only be realized in an open quantum system that interacts with the environment causing dephasing or decoherence, captured by the Lindblad equation which we discuss next. A previous study considered a similar time evolution for only dislocation modes but in a closed system~\cite{OpenTopo:10}. The results reported therein, however, remains qualitatively valid for sufficiently weak system-to-environment coupling.

\subsection{Quantum dynamics in open systems}~\label{subsec:lindbald}

Decoherence in quantum systems is a vast and rich subject that has been discussed in literature for decades~\cite{LE:1, LE:2, LE:3, LE:4, LE:5, LE:6, LE:7, LE:8, LE:9, LE:10, LE:11, LE:12, LE:13, LE:14}. Here we model the decoherence or dephasing in open quantum systems such that the environment monitors the energy of the quantum system through continuous measurements. Under this assumption, the temporal dynamics of the quantum system is non-unitary that is, however, smoothly connected to the unitary time evolution in closed quantum systems. The derivation of the corresponding Lindblad equation has been detailed in Ref.~\cite{OpenTopo:9}. We here briefly review the assumptions leading to a specific form of the Linblad equation displayed below in Eq.~\eqref{eq:lindbald}. In this formulation, the quantum system of interest is made to evolve in conjunction with an ancillary system connected to the environment, such that they are coupled together via some interaction Hamiltonian. The whole system is allowed to jointly evolve with time while projective measurements are made only on the ancilla. This procedure effectively provides a quantum channel for monitoring the energy of quantum systems by the environment leading to decoherence of the system in the energy basis. Finally, assuming that there is no memory effect in the environment, such that the dynamics is Markovian, we arrive at the master or Lindblad equation
\begin{equation}~\label{eq:lindbald}
\frac{\partial \rho(t)}{\partial t}= -i \left[H(t),\rho(t) \right] - \gamma \left[ H(t), \left[ H(t), \rho(t)\right] \right]
\end{equation} 
in which the Lindblad operator is the time-dependent Hamiltonian $H(t)$, obtained upon setting the reduced Planck constant $\hbar=1$. Here, $\rho(t)$ is the time-dependent density matrix of the open quantum system. In this situation, the open quantum system decoheres among the energy eigenstates of the instantaneous Hamiltonian $H(t)$. The parameter $\gamma$ determines the decoherence rate as well as the strength of measurement, yielding non-unitary time dynamics of an open quantum system. For $\gamma=0$ we recover the unitary time dynamics of a closed quantum system. We numerically solve Eq.~\eqref{eq:lindbald} to unfold the time dynamics of various topologically protected modes in open topological crystals. The structures of the associated density matrices are discussed in the next subsection.

\begin{figure}[t!]
\includegraphics[width=1.00\linewidth]{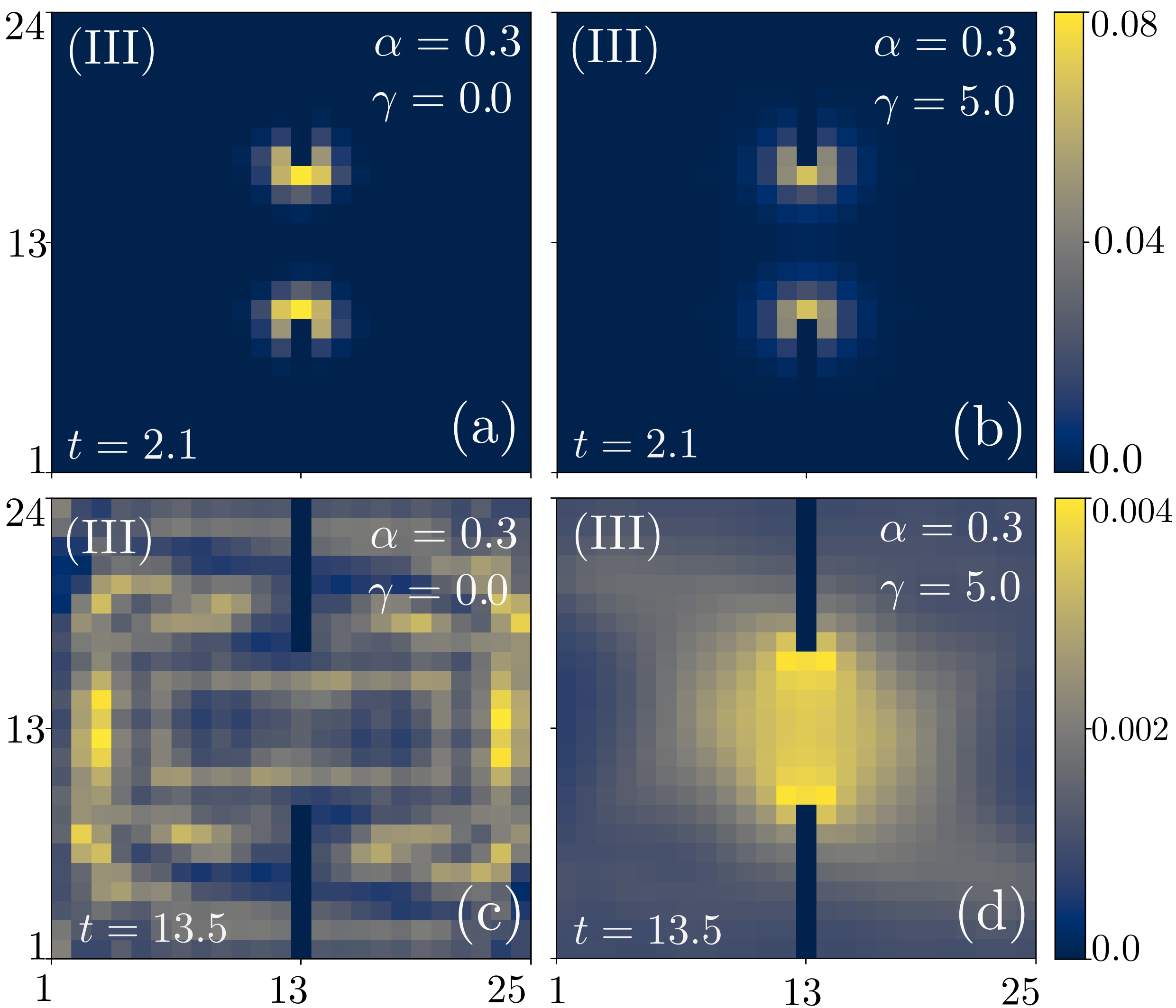}
\caption{Same as Fig.~\ref{fig:DislocationMeltingLDOS1}, but for $m_i=-1.0$ and $m_f=-2.5$. The corresponding time evolution of the probability of finding the dislocation modes at any time $t$ is shown in Fig.~\ref{fig:DislocationMeltingProb}(e). The Roman numeral in each panel corresponds to the arrow out of the ${\rm M}$ phase, shown in Fig.~\ref{fig:SetupDislocation}(c).     
}~\label{fig:DislocationMeltingLDOS3}
\end{figure}

\begin{figure}[t!]
\includegraphics[width=1.00\linewidth]{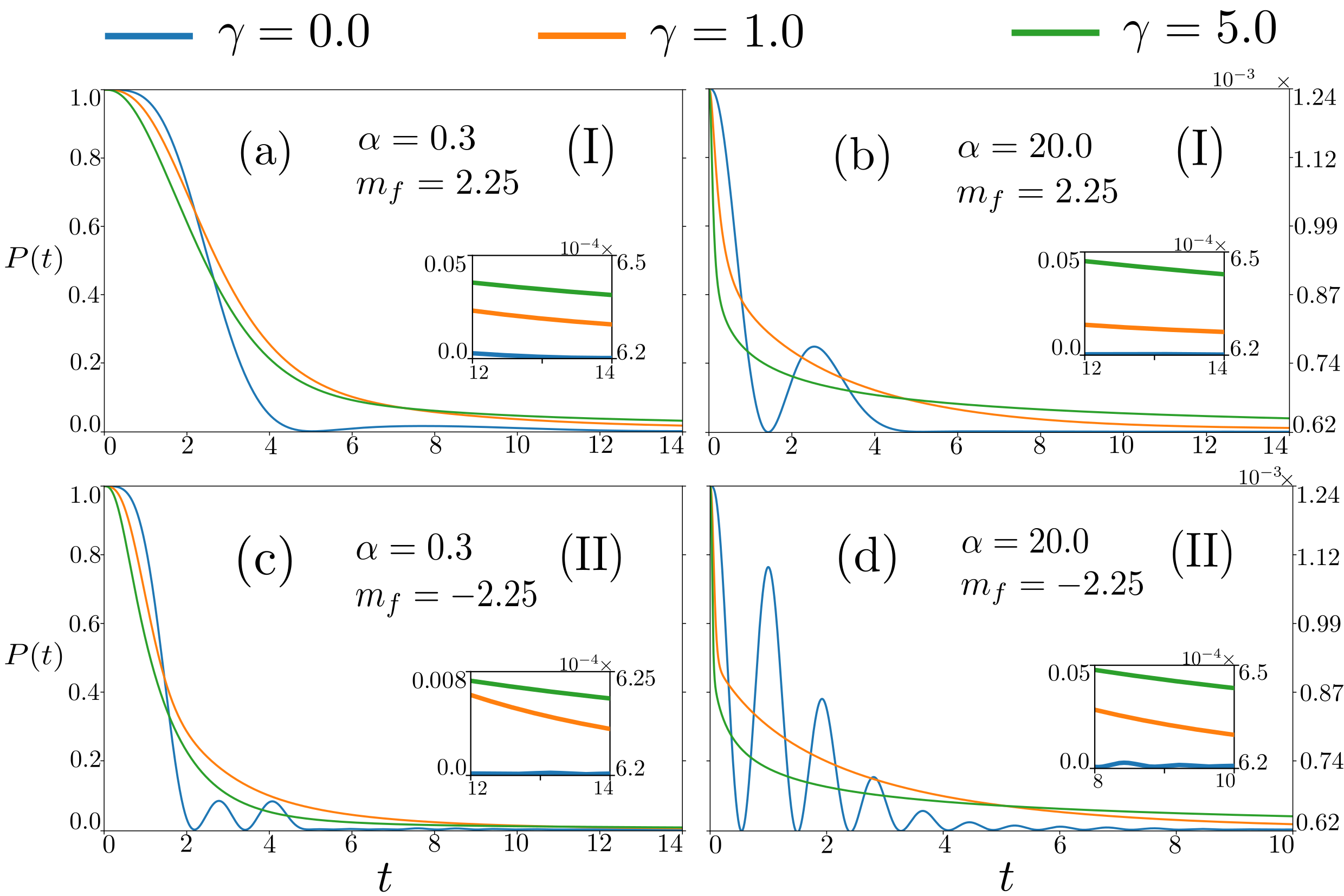}
\caption{Probability $P(t)$ of finding the corner modes [Eq.~\eqref{eq:prob}] in the presence of a real time ramp that takes the system initially in second-order  topological insulator (SOTI) phase with $m_i=1$ at $t=0$ to a normal insulator with (a) and (b) $m_f=2.25$, and (c) and (d) $m_f=-2.25$ for a slow ramp with $\alpha=0.3$ [(a) and (c)] and a fast ramp or a sudden quench with $\alpha=20$ [(b) and (d)] for various choices of $\gamma$ quantifying the system-to-environment coupling. See Eqs.~\eqref{eq:ramprofile} and~\eqref{eq:lindbald}. We compute such probability from a pure state (with the numbers shown on the left vertical axis) and a mixed state ${\rm HF}^{\prime \prime}$ (with the numbers shown on the right vertical axis). Inset in each subfigure shows the long-time behavior of $P(t)$ for various $\gamma$ values. The real time ramp begins at $t=0$ and thus we show $P(t)$ for $t>0$ only. For $\gamma=0$ we recover the results for the unitary time evolution in a closed system. The Roman numeral in each panel corresponds to the arrow out of the SOTI phase, shown in Fig.~\ref{fig:SetupCorner}(c). Here, we set $\Delta=1.0$.  
}~\label{fig:CornerMeltingProb}
\end{figure}

\subsection{Density matrix}~\label{subsec:densitymatrix}

In this work, we are interested in two different types of time dynamics in open topological systems. First, we consider a situation in which the initial system at $t=0$ supports topological edge or dislocation or corner modes. But, under the time evolution the system finally ends up in an insulating phase via a continuous quantum phase transition that is devoid of any such topological modes. Then, we seek to examine the fate of these topological modes under the non-unitary time evolution governed by Eq.~\eqref{eq:lindbald}. In this case, we can consider a simpler situation when the initial density matrix represents a \emph{pure} state, constituted by one of the edge ($\eta={\rm edge}$) or dislocation ($\eta={\rm dis}$) or corner ($\eta={\rm cor}$) modes, and is explicitly given by 
\begin{equation}~\label{eq:DMpure}
\rho^{\eta}_j(0)= \ket{\Psi^{\eta}_j} \bra{\Psi^{\eta}_j},
\end{equation} 
where $j$ can take value 1 or 2 for the edge and dislocation modes, and 1 or 2 or 3 or 4 for the corner mode. Then, once the time-dependent ramp [Eq.~\eqref{eq:ramprofile}] is switched on, at any instant of time ($t$) the probability of finding the topological mode is given by 
\begin{equation}~\label{eq:prob}
P(t)= \bra{\Psi^{\eta}_j} \rho(t) \ket{\Psi^{\eta}_j},
\end{equation}
which is also known as the \emph{fidelity}~\cite{fidelity:1, fidelity:2}. In the last equation $\rho(t) \equiv \rho^{\eta}_j(t)$, which we obtain by numerically solving Eq.~\eqref{eq:lindbald} for various choices of $m_i$, $m_f$, $\alpha$, and $\gamma$, as discussed in the next section.

\begin{figure}[t!]
\includegraphics[width=1.00\linewidth]{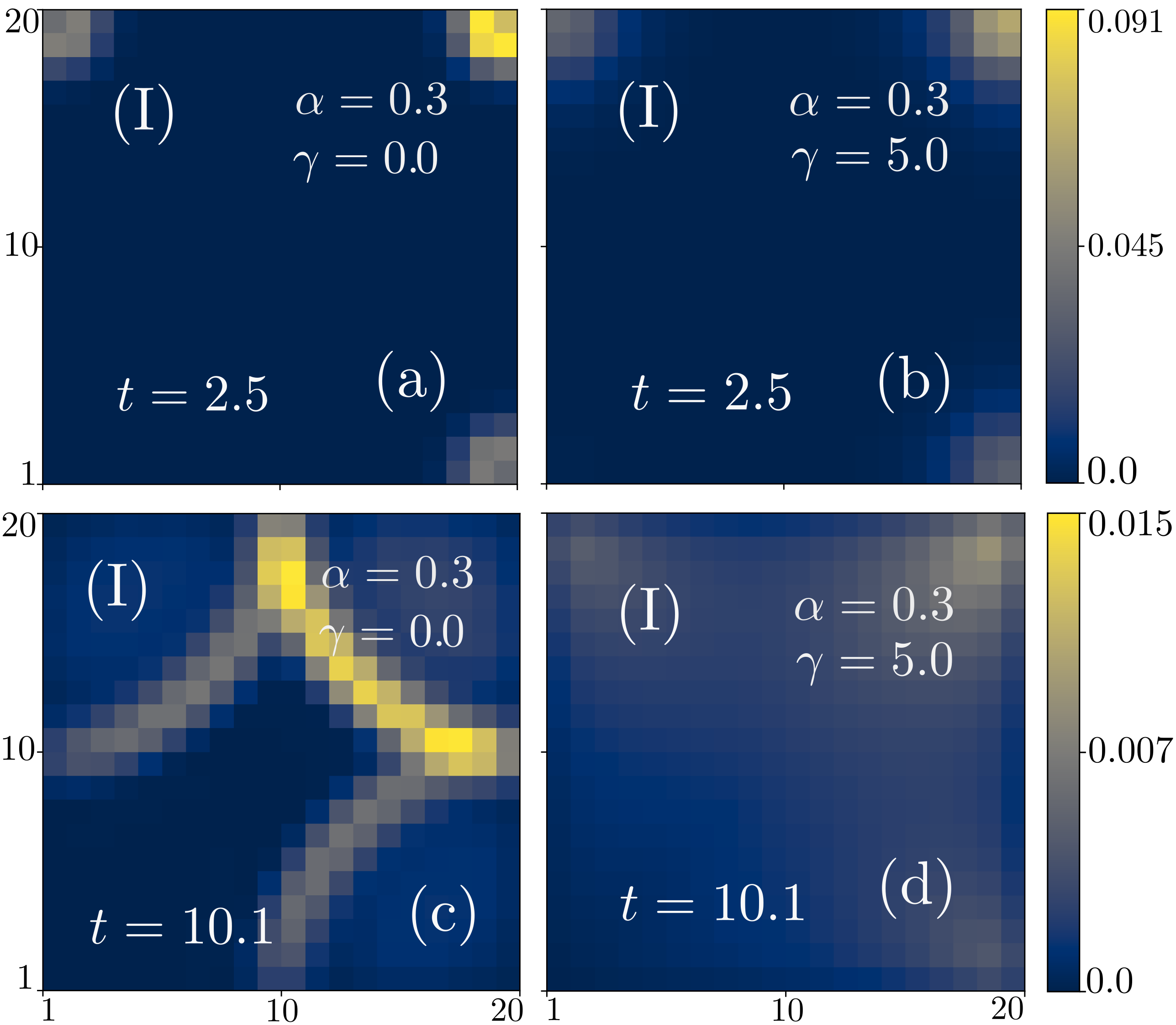}
\caption{Site-resolved local density of states (LDOS), computed from the density matrix $\rho(t)$ [see Eq.~\eqref{eq:LDOS}] when the initial density matrix $\rho(0)$ is constructed from a single corner mode (pure state) for a slow ramp with $\alpha=0.3$ at (a) and (b) an earlier time instant $t=2.5$ and (c) and (d) a large time instant $t=10.1$. The left panels [(a) and (c)] correspond to a unitary time evolution in a closed system with $\gamma=0$ while the right panels [(b) and (d)] correspond to a non-unitary time evolution in an open system with the system-to-environment coupling strength $\gamma=5.0$. See Eqs.~\eqref{eq:ramprofile} and~\eqref{eq:lindbald}. Here we choose $m_i=1.0$, $m_f=2.25$, and $\Delta=1.0$. We arrive at the same profile for the site-resolved LDOS when computed from a mixed state ${\rm HF}^{\prime \prime}$ once the uniform background LDOS for the half-filled system is subtracted. The corresponding time evolution of the probability of finding the corner modes at any time $t$ is shown in Fig.~\ref{fig:CornerMeltingProb}(a). The Roman numeral in each panel corresponds to the arrow out of the second-order topological insulator phase, shown in Fig.~\ref{fig:SetupCorner}(c).  
}~\label{fig:CornerMeltingLDOS1}
\end{figure}

From the solutions of $\rho(t)$ we can also extract the site-resolved LDOS for the topological modes at any instant of time $t$, given by 
\begin{equation}~\label{eq:LDOS}
D_i (t) = \sum_{\beta \in {\rm IDF}} \bra{i,\beta} \rho(t) \ket{i,\beta},
\end{equation} 
where $i$ is the site index, and the summation over $\beta$ runs over all the appropriate internal degrees of freedom (IDF), such as orbital and spin. The single-particle state at site $i$ with IDF $\beta$ is denoted by $\ket{i,\beta}$. In the next section, we also discuss the time evolution of the LDOS for various topological modes.

However, in topological quantum materials, robust midgap modes can only be occupied upon filling all the bulk states at negative energies. Furthermore, in a half-filled system, the electronic LDOS is uniform throughout the whole system due to the unitary or anti-unitary particle-hole symmetry, discussed in the previous section, irrespective of whether the system is in the topological or trivial parameter regime. Thus, in order to relate our findings on the temporal dynamics of topological modes in open quantum systems, we need to consider an appropriate many-body ground state at the initial time $t=0$. Namely we need to add one (two) fermion(s) to the half-filled system, a state denoted by ${\rm HF}^{\prime}$ (${\rm HF}^{\prime \prime}$) while tracking the time dynamics of the edge and dislocation (corner) modes. When the topological modes exist, in the ${\rm HF}^{\prime}$ and ${\rm HF}^{\prime \prime}$ states the LDOS shows peaks where such modes are localized once the background uniform LDOS of the half-filled system is subtracted. The corresponding initial \emph{mixed} state at $t=0$ in a system with $N$ number of lattice sites reads as 
\begin{equation}~\label{eq:mixed}
\rho(0)=\frac{1}{j(N+1)} \; \sum^{j(N+1)}_{i=1} \ket{\Psi_i} \bra{\Psi_i}
\end{equation}   
where $j=1$ for the edge and dislocation modes and $j=2$ for the corner modes, and $\ket{\Psi_i}$ is the eigenstate of the corresponding initial Hamiltonian $H_j(0)$. Notice that in ${\rm HF}^{\prime}$ state two edge and dislocation modes are occupied, besides $N-1$ number of bulk modes. In ${\rm HF}^{\prime \prime}$ state $N-2$ bulk modes and four corner modes are occupied.

The time evolution of the mixed state density matrix from Eq.~\eqref{eq:mixed} is also obtained by numerically solving Eq.~\eqref{eq:lindbald}. From these solutions, we can obtain the probability of finding the edge, dislocation, and corner modes at any $t \geq 0$ from Eq.~\eqref{eq:prob} after replacing $\ket{\Psi^{\eta}_j}$ by  
\begin{equation}~\label{eq:wavefunctionmixed}
\ket{\Psi}= \sum^{2}_{j=1} \frac{\ket{\Psi^{\rm edge}_j}}{\sqrt{2}}, \:\:\:
\sum^{2}_{j=1} \frac{\ket{\Psi^{\rm dis}_j}}{\sqrt{2}}, \:\: \text{and} \:\:\:
\sum^{4}_{j=1} \frac{\ket{\Psi^{\rm cor}_j}}{\sqrt{4}},
\end{equation}
respectively, therein. From the numerical solution of $\rho(t)$, now obtained with $\rho(0)$ from Eq.~\eqref{eq:mixed} as the initial condition, we can also obtain the site-resolved LDOS upon replacing $\ket{\Psi^{\eta}_j}$ by $\ket{\Psi}$ in Eq.~\eqref{eq:LDOS}, which we compute after subtracting the background uniform LDOS in the half-filled system. Then the results at $t=0$ are identical to the ones shown in Fig.~\ref{fig:SetupEdge}(b),~\ref{fig:SetupDislocation}(b), and~\ref{fig:SetupCorner}(b).

\begin{figure}[t!]
\includegraphics[width=1.00\linewidth]{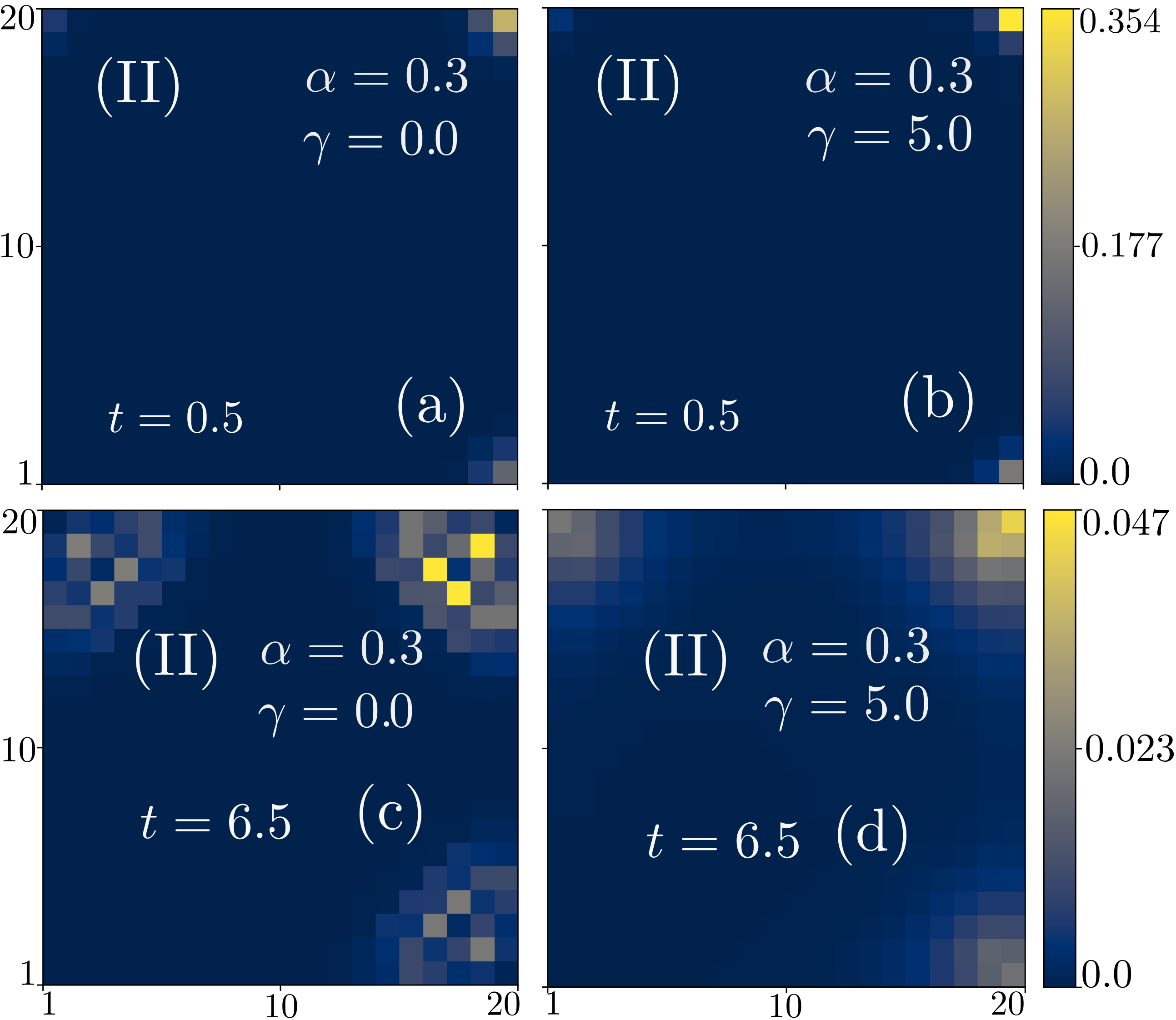}
\caption{Same as Fig.~\ref{fig:CornerMeltingLDOS1}, but for $m_i=1.0$, $m_f=-2.25$, and $\Delta=1.0$. The corresponding time evolution of the probability of finding the corner modes at any time $t$ is shown in Fig.~\ref{fig:CornerMeltingProb}(c). The Roman numeral in each panel corresponds to the arrow out of the second-order topological insulator phase, see Fig.~\ref{fig:SetupCorner}(c). Notice that at $t=0.5$ the site-resolved local density of states (LDOS) in a closed system with $\gamma=0$, see panel (a), is predominantly localized around a few sites near the top-right corner with comparable intensities, while that in an open system with $\gamma=5$, see panel (b), shows a peak only at a single site at the top-left corner. Therefore, although the LDOS at the top-right site is stronger for $\gamma=5$ in comparison to that when $\gamma=0$, the overlap with the initial corner mode [see Fig.~\ref{fig:SetupCorner}(b)], $P(t=0.5)$ is larger in a closed system.      
}~\label{fig:CornerMeltingLDOS2}
\end{figure}

In the next part of this quest, we seek to demonstrate the dynamic condensation of these topological modes, when the real time ramp begins in an insulating phase, devoid of such modes but ends in an insulator that otherwise fosters such topologically robust gapless modes at least in the static system. As in this scenario, there are no topological modes to begin with at $t=0$, our analysis always needs to be performed with a mixed state, for which the density matrix is shown in Eq.~\eqref{eq:mixed}. In this case, the initial state HF$^\prime$ is devoid of any topological mode or does not possess the ones supported by the final state, which is the case only for the ramp (IV) shown in Fig.~\ref{fig:SetupEdge}(c), while HF$^{\prime \prime}$ is always devoid of any topological corner modes. The rest of the procedure is already discussed in the last three paragraphs, which also applies here to underpin the dynamic condensation of the edge, dislocation, and corner modes. Now the stage is set with the discussion on all the requisite formalism in order to promote dynamic melting and condensation of these topological modes in square lattice-based open quantum systems in two spatial dimensions.

\section{Time dynamics: Results}~\label{sec:Tdynamicsresults}

\begin{figure*}[t!]
\includegraphics[width=1.00\linewidth]{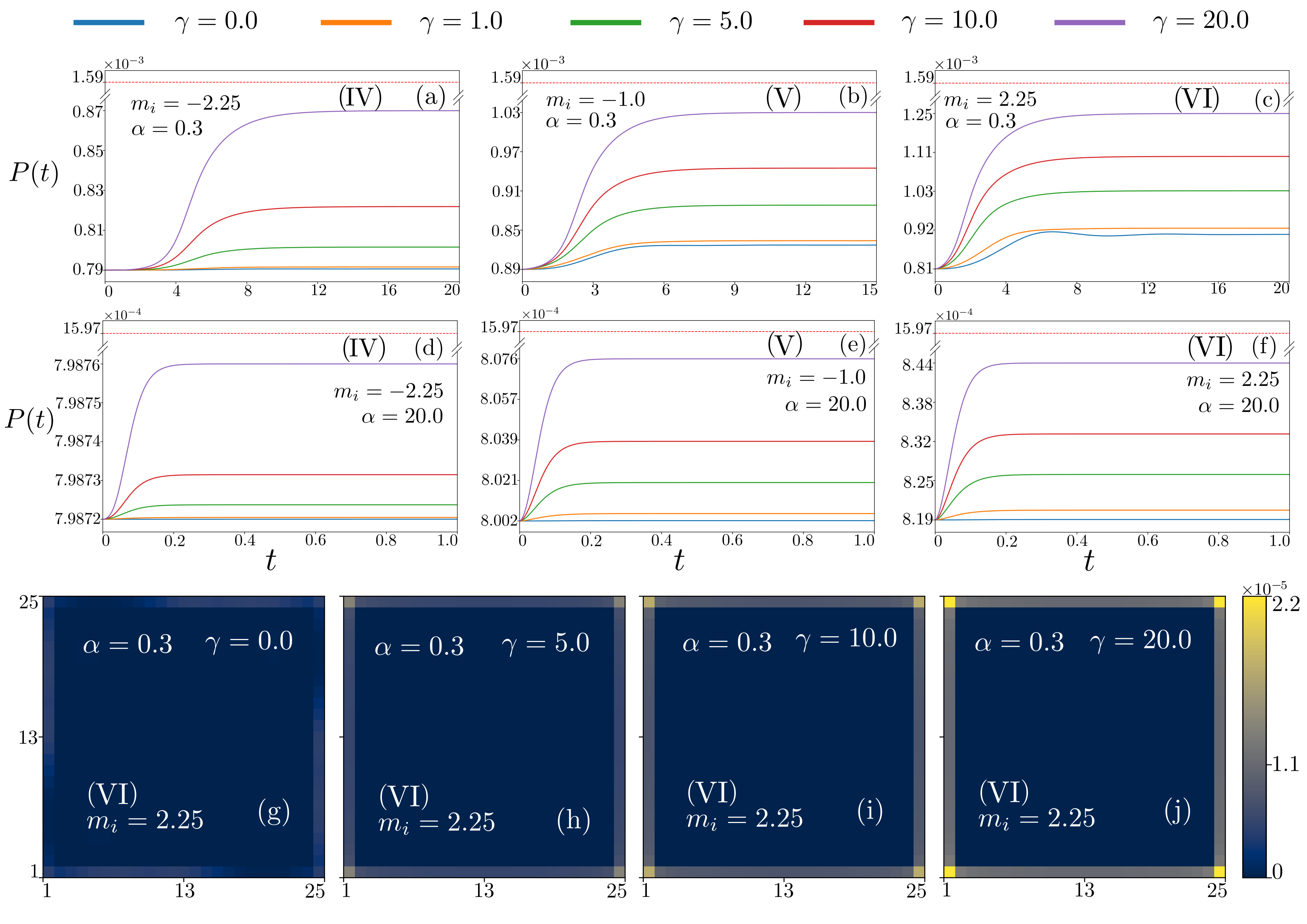}
\caption{Probability $P(t)$ of dynamic condensation of the edge modes [Eq.~\eqref{eq:prob}] for (a) and (d) $m_i=-2.25$, (b) and (e) $m_i=-1.0$, and (c) and (f) $m_i=2.25$ when $m_f=1.0$. The top row [(a)-(c)] corresponds to a slow ramp with $\alpha=0.3$ and the second row [(d)-(f)] corresponds to a fast ramp or a sudden quench with $\alpha=20$. In each panel, plots corresponding to various choices of the system-to-environment coupling strength $\gamma$ are shown in different colors. See Eqs.~\eqref{eq:ramprofile} and~\eqref{eq:lindbald}. For $\gamma=0$ we recover the results under the unitary time evolution in a closed system. Dashed horizontal line in each panel depicts the maximal attainable probability value in a closed system. Results are obtained for a mixed state ${\rm HF}^{\prime}$. The site-resolved saturated local density of states (LDOS), see Eq.~\eqref{eq:LDOS}, for $m_i=2.25$, $m_f=1$, $\alpha=0.3$, and (g) $\gamma=0$ (closed system), (h) $\gamma=5.0$, (i) $\gamma=10.0$, and (j) $\gamma=20.0$, show an increasing localization at the edges with increasing system-to-environment coupling ($\gamma$). Here we display LDOS after subtracting its uniform background value in the half-filled system. The Roman numeral in each panel corresponds to the arrow into the $\Gamma$ phase, shown in Fig.~\ref{fig:SetupEdge}(c).   
}~\label{fig:EdgeCondensation}
\end{figure*}

\begin{figure*}[t!]
\includegraphics[width=1.00\linewidth]{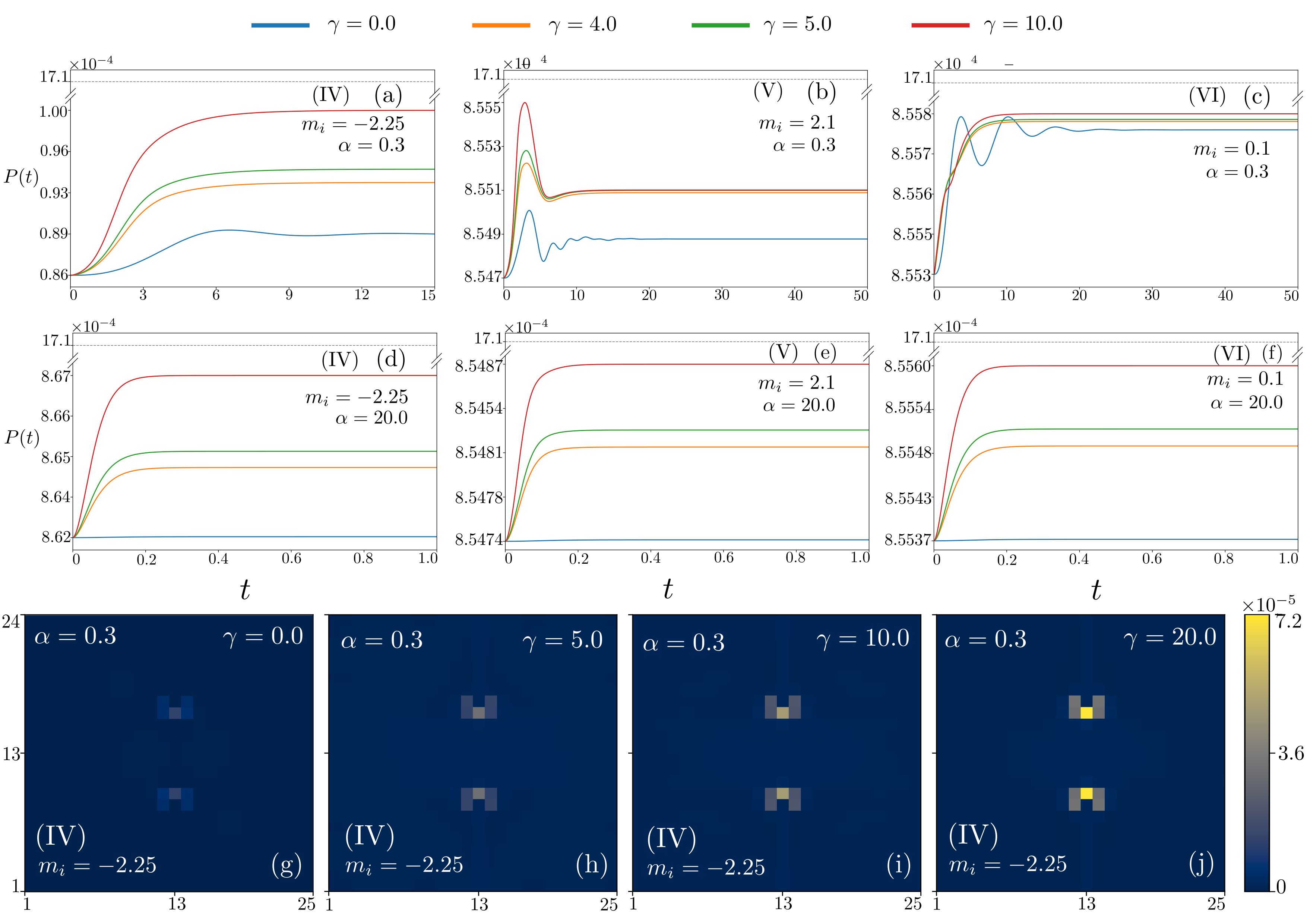}
\caption{Probability $P(t)$ of dynamic condensation of the dislocation modes [Eq.~\eqref{eq:prob}] for (a) and (d) $m_i=-2.25$, (b) and (e) $m_i=2.1$, and (c) and (f) $m_i=0.1$ when $m_f=-1.0$. The top row [(a)-(c)] corresponds to a slow ramp with $\alpha=0.3$ and the second row [(d)-(f)] corresponds to a fast ramp or a sudden quench with $\alpha=20$. In each panel, plots corresponding to various choices of the system-to-environment coupling strength $\gamma$ are shown in different colors. See Eqs.~\eqref{eq:ramprofile} and~\eqref{eq:lindbald}. For $\gamma=0$ we recover the results under the unitary time evolution in a closed system. Dashed horizontal line in each panel depicts the maximal attainable probability value in a closed system. Results are obtained for a mixed state ${\rm HF}^{\prime}$. The site-resolved saturated local density of states (LDOS), see Eq.~\eqref{eq:LDOS}, for $m_i=-2.25$, $m_f=1$, $\alpha=0.3$, and (g) $\gamma=0$ (closed system), (h) $\gamma=5.0$, (i) $\gamma=10.0$, and (j) $\gamma=20.0$, show an increasing localization at the cores of dislocation and antidislocation with increasing system-to-environment coupling ($\gamma$). Here we display LDOS after subtracting its uniform background value in the half-filled system. The Roman numeral in each panel corresponds to the arrow into the translationally active ${\rm M}$ phase, see Fig.~\ref{fig:SetupDislocation}(c).   
}~\label{fig:DislocationCondensation}
\end{figure*}

\begin{figure*}[t!]
\includegraphics[width=0.80\linewidth]{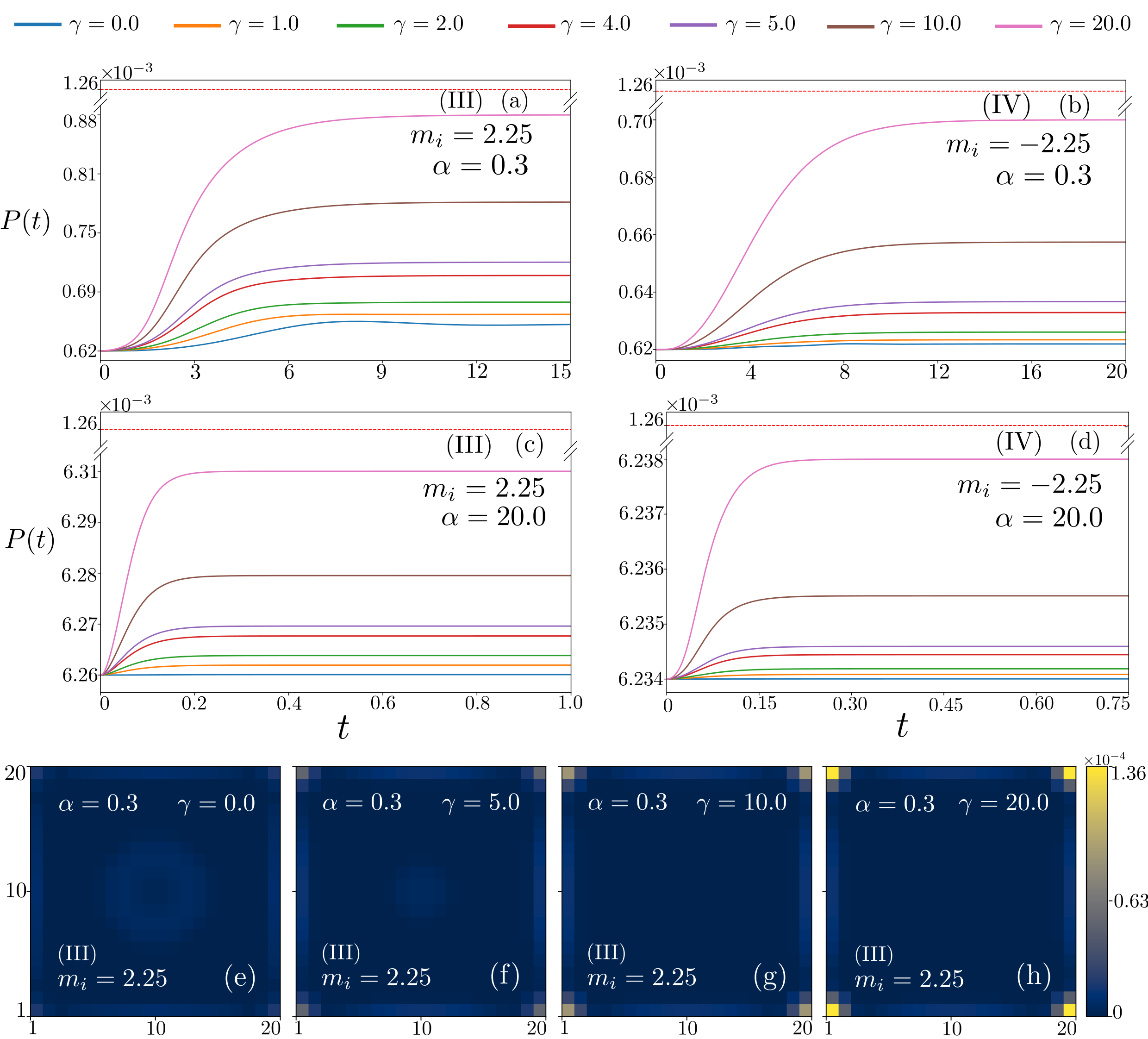}
\caption{Probability $P(t)$ of dynamic condensation of the corner modes [Eq.~\eqref{eq:prob}] for (a) and (c) $m_i=2.25$, and (b) and (d) $m_i=-2.25$, when $m_f=1.0$ and $\Delta=1.0$. The top row [(a) and (b)] corresponds to a slow ramp with $\alpha=0.3$ and the second row [(c) and (d)] corresponds to a fast ramp or a sudden quench with $\alpha=20$. In each panel, plots corresponding to various choices of the system-to-environment coupling strength $\gamma$ are shown in different colors. See Eqs.~\eqref{eq:ramprofile} and~\eqref{eq:lindbald}. For $\gamma=0$ we recover the results for unitary time evolution in a closed system. Dashed horizontal line in each panel depicts the maximal attainable probability value in a closed system. Results are obtained for a mixed state ${\rm HF}^{\prime \prime}$. Site-resolved saturated local density of states (LDOS), see Eq.~\eqref{eq:LDOS}, for $m_i=2.25$, $m_f=1$, $\Delta=1.0$, $\alpha=0.3$, and (e) $\gamma=0$ (closed system), (f) $\gamma=5.0$, (g) $\gamma=10.0$, and (h) $\gamma=20.0$, show an increasing localization at the corners of a square lattice with increasing system-to-environment coupling ($\gamma$). Here we display LDOS after subtracting its uniform background value in the half-filled system. The Roman numeral in each panel corresponds to the arrow into the second-order topological insulator phase, shown in Fig.~\ref{fig:SetupCorner}(c).
}~\label{fig:CornerCondensation}
\end{figure*}

In this section, we discuss the dynamic melting and condensation of topological edge, dislocation, and corner modes employing the theoretical tools discussed in the previous section. From the numerical solutions of the Lindblad equation [see Eq.~\eqref{eq:lindbald}] with a time-dependent mass or Hamiltonian [see Eq.~\eqref{eq:ramprofile}], we note that the dynamic melting for all types of topological modes are qualitatively similar (see Figs.~~\ref{fig:EdgeMeltingProb}-\ref{fig:CornerMeltingLDOS2}), and the same is true for their dynamic condensation (see Figs.~\ref{fig:EdgeCondensation}-\ref{fig:CornerCondensation}). Therefore, we discuss these two phenomena for all types of topological modes in two subsequent subsections.

\subsection{Dynamic melting}~\label{subsec:Tdynamicsresultsmelting}

First we consider a situation when the real time ramp begins in an insulating state that accommodates edge or dislocation or corner modes, and ends in a gapped state where such modes are absent. Any such time evolution always takes the system through at least one quantum critical point, where the bulk gap $\Delta_{\rm bulk}$ closes. Although in topological phases, there exist gapless modes, they are much fewer in number in comparison to the number of bulk states. Therefore, the dynamics of the system is always controlled by the minimal energy gap between the filled and empty bulk states at positive and negative energies, respectively. At the beginning of a real time ramp when the system is far from such critical point, the topological modes decohere among all the eigenstates of the instantaneous Hamiltonian. With increasing system-to-environment coupling ($\gamma$) the strength of such decoherence increases, yielding a faster drop in the probability of finding such modes $P(t)$ with increasing $\gamma$. This process continues until the system arrives at the shore of a quantum phase transition where $\Delta_{\rm bulk} \to 0$ continuously. Then the system suffers a critical slow down due to the divergence of the associated time scales therein, which vary inversely with the energy gap ($\Delta_{\rm bulk}$). In this regime, the system falls out of equilibrium, and the dynamics of the system is determined by the competition between two time scales, namely the correlation time ($\tau_{\rm cor}$) and the decoherence time ($\tau_{\rm dec}$), given by~\cite{OpenTopo:9} 
\allowdisplaybreaks[4]
\begin{eqnarray}
&& \tau_{\rm cor} \sim \frac{1}{\Delta_{\rm bulk}} \sim \left( \alpha |t-t_0| \right)^{-\nu z} \nonumber \\
&& \text{and} \:\:\:
\tau_{\rm dec} \sim \frac{1}{\gamma \Delta^2_{\rm bulk}} \sim \frac{\left( \alpha |t-t_0| \right)^{-2\nu z}}{\gamma},
\end{eqnarray}
respectively, with $z=\nu=1$ for all the models we discuss in this work. The band gap closes at time $t=t_0$.

The freeze-out time $\tau_{\rm freeze}$ of the critical system is determined by ${\rm min}.(\tau_{\rm cor},\tau_{\rm dec})$. When $\tau_{\rm freeze}=\tau_{\rm cor}$, the time evolution of the system remains unitary, as is the case for $\gamma=0$ (closed system) and sufficiently small $\gamma \; (\ll 1)$. On the other hand, when $\tau_{\rm freeze}=\tau_{\rm dec}$ the system exhibits the quantum Zeno effect in which it gets arrested in a particular configuration as repeated measurements slow down the evolution and eventual decay of the survival probability with time. In other words, the system then does not appear to change due to frequent measurements. As near the critical point, $\gamma \sim \alpha^{\nu z/(1+\nu z)}$~\cite{OpenTopo:9}, for sufficiently strong system-to-environment coupling $\gamma \gtrsim 1$ (holds for all our calculations) the freeze-out time is determined by the decoherence time, and the quantum Zeno effect governs the time dynamics of the system beyond the quantum phase transition into the final state. Consequently, the decay of the survival probability $P(t)$ is slowed down to a larger extent due to the Quantum Zeno effect leading to a greater saturation values (although small) for greater measurement strength $\gamma$. Finally, however, $P(t) \to 0$ as $t \to \infty$ slowly, as the final insulating state does not harbor the topological modes of the initial state. Furthermore, with increasing ramp speed ($\alpha$) the system arrives at the brink of topological quantum phase transition sooner, and then the quantum Zeno effect sets in at an earlier time. These outcomes are displayed in Figs.~\ref{fig:EdgeMeltingProb},~\ref{fig:DislocationMeltingProb}, and~\ref{fig:CornerMeltingProb} for the edge, dislocation, and corner modes for various choices of the real time ramp.

Notice that in a closed quantum system $\tau_{\rm dec} \to \infty$ as $\gamma \to 0$ therein, and the quantum Zeno effect never develops in the system. As a result, $P(t)$ after displaying oscillatory behavior ultimately vanishes. These features are also displayed in Figs.~\ref{fig:EdgeMeltingProb},~\ref{fig:DislocationMeltingProb}, and~\ref{fig:CornerMeltingProb}. We also note that for $\gamma \gtrsim 1$, the temporal behavior of $P(t)$ is devoid of any oscillatory behavior due to strong dephasing at short and quantum Zeno effect in the long time scales. Nevertheless, as $\gamma \to 0$, the system slowly starts to develop the oscillatory behavior in $P(t)$ as is the case in closed system ($\gamma=0$) as then $\tau_{\rm cor}$ sets the freezing time. We, however, we do not show it explicitly in this work.

We now discuss the imprints of the time evolution of $P(t)$ on the site-resolved LDOS. At a short time scale (long before the system undergoes a quantum phase transition), the LDOS is highly localized at the edges, cores of the pair of dislocation and anti-dislocation, and corners in a first-order TI, a translationally active first-order TI, and a SOTI, respectively. But, the intensity of the associated LDOS gets weaker with an increasing system-to-environment coupling, as can be seen by comparing the LDOS for $\gamma=0$ and $\gamma=5$ for a fixed ramp speed $\alpha=0.3$, for example. The results are shown in panels (a) and (b), respectively, of Figs.~\ref{fig:EdgeMeltingLDOS1}-\ref{fig:EdgeMeltingLDOS3} (for edge modes), Figs.~\ref{fig:DislocationMeltingLDOS1}-\ref{fig:DislocationMeltingLDOS3} (for dislocation modes), and Figs.~\ref{fig:CornerMeltingLDOS1} and~\ref{fig:CornerMeltingLDOS2} (for corner modes). This behavior is naturally insensitive (qualitatively) to the nature of the final state, and solely attributed to strong decoherence of the initial state among all the energy eigenstates of the instantaneous Hamiltonian at a given instant of time.

At a later time (after the system encounters a quantum phase transition through band gap closing), the situation with the site-resolved LDOS gets reversed as then the quantum Zeno effect determines the time dynamics of the system for large $\gamma$, otherwise absent in a closed system ($\gamma=0$). In a closed system we find that the LDOS spreads over the entire system with a few peaks that are placed away from the edges or defect cores or corners conforming with the fact that $P(t)$ drops appreciably close to zero by then. By contrast, in an open system with strong system-to-environment coupling, as is the case when we set $\gamma=5$, the LDOS continues to peak around the edges or defect cores or corners, which are nonetheless weaker than the ones we find at an earlier time prior to the topological quantum phase transition. These features in LDOS are consistent with the variation of $P(t)$ at long time scale where it gets larger with increasing $\gamma$ due to the quantum Zeno effect. The representative results for $\gamma=0$ and $\gamma=5$ are respectively shown in panels (c) and (d) of Figs.~\ref{fig:EdgeMeltingLDOS1}-\ref{fig:EdgeMeltingLDOS3} (for edge modes), Figs.~\ref{fig:DislocationMeltingLDOS1}-\ref{fig:DislocationMeltingLDOS3} (for dislocation modes), and Figs.~\ref{fig:CornerMeltingLDOS1} and~\ref{fig:CornerMeltingLDOS2} (for corner modes) for a fixed ramp speed $\alpha=0.3$.

As a final remark on dynamic melting of topological modes, we point out that the reported profiles of LDOS are the same when computed from a pure state and from a mixed state (HF$^\prime$ or HF$^{\prime \prime}$). But, upon computing the LDOS from a mixed state we always subtract its uniform background value in the half-filled system.

\subsection{Dynamic condensation}~\label{subsec:Tdynamicsresultscondensation}

Next, we consider a reverse course of the time evolution in which the system always ends up (as $t \to \infty$) in an insulating phase that supports edge or dislocation or corner modes. But, the real time ramp begins in a gapped state at $t=0$ that is either devoid of any such mode or supports topological modes that are distinct from the ones we find in the final state of matter. The latter is pertinent only when we consider a real time ramp from the ${\rm M}$ phase to the ${\rm \Gamma}$ phase [see ramp (IV) in Fig.~\ref{fig:SetupEdge}(c)], both supporting edge modes which are, however, distinct.

Irrespective of these details, we find that in a closed quantum system ($\gamma=0$) the condensation probability $P(t)$ of such topological modes always increases with decreasing ramp speed ($\alpha$). This observation conforms with the adiabatic theorem, indicating that for sufficiently slow time evolution, the energy levels get mapped to corresponding instantaneous eigenstates as the system evolves with time, which in the final state (as $t \to \infty$) harbors the topological mode because of the way the initial mixed density matrix (for HF$^{\prime}$ and HF$^{\prime \prime}$) is prepared. More importantly, we find that for any fixed ramp speed, the condensation probability of any topological mode increases with increasing system-to-environment coupling ($\gamma$) as $t \to \infty$. This phenomenon can be explained in the following way. With increasing $\gamma$, the degree of decoherence of the initial state among all the energy eigenstates of the final Hamiltonian that also fosters desired topological mid-gap modes, increases, thereby yielding a larger $P(t \to \infty)$ for larger $\gamma$ values. Furthermore, we also notice that for any fixed but non-trivial $\gamma$, $P(t \to \infty)$ increases with the decreasing ramp speed $\alpha$, possibly indicating an extended jurisdiction of the adiabatic theorem even in open quantum systems. We arrive at these conclusions from the complete time evolution of $P(t)$ and the profile of the saturated (time-independent) site-resolved LDOS in the final state. The explicit results are shown in Fig.~\ref{fig:EdgeCondensation} (for edge modes), Fig.~\ref{fig:DislocationCondensation} (for dislocation modes), and Fig.~\ref{fig:CornerCondensation} (for corner modes).

\section{Summary and discussions}~\label{sec:summary}

To summarize, here we investigate the non-unitary time dynamics of topological edge, dislocation, and corner modes, hallmarks of various 2D TIs, when the underlying lattice-based open quantum system evolves between topologically distinct insulating states in real time and couples to the environment through energy channels, thereby sourcing decoherence in the system without generating high energy excitations. Primarily, we focus on a situation in which the initial state at time $t=0$ is endowed with such topological mid-gap modes, living close to the zero-energy, while the final state, reached via a real time ramp of the mass parameter as $t \to \infty$, is devoid of such modes. In that situation, we find that with increasing system-to-environment coupling ($\gamma$) the signature of these modes decreases more rapidly at small time scale before the system undergoes a topological quantum phase transition, identified by a bulk band gap closing. This phenomenon is credited to the strong decoherence of the initial state among all the energy eigenstates of the time-evolving instantaneous Hamiltonian. But, in the long time scale (after the quantum phase transition when the bulk gap reopens) the quantum Zeno effect sets in and the system then effectively gets stuck in a particular configuration as repeated measurements slow down its evolution with time. Consequently, in this regime the probability of finding the initial topological mode (although small) then increases with increasing $\gamma$. On the other hand, for a reverse course of the time evolution the saturated (with respect to time) condensation probability of such localized topological zero-energy modes, harbored by the final state, is shown to increase with the increasing system-to-environment coupling at the end of the real time ramp. This outcome possibly results from the fact that with increasing $\gamma$, the initial state decoheres among all the eigenstates of the final Hamiltonian more strongly which also include the topological modes. We establish these outcomes from the complete time evolution of the probability of finding such modes $P(t)$ and the site-resolved LDOS at various time instants. The results are summarized in Figs.~\ref{fig:EdgeMeltingProb}-~\ref{fig:CornerCondensation}.

Our formalism, based on density matrix and Lindblad equation, can be applied to study non-unitary time dynamics of all types of topological modes realized in any open topological quantum crystals of arbitrary dimensionality and symmetry class. And the results therein should be qualitatively similar to the ones reported in the present work. In future, these avenues can be explored systematically. Furthermore, in future it will be worthwhile to scrutinize the fate of these modes under different types of system-to-environment coupling. Among various possibilities, the situation in which the topological quantum system is coupled to an external thermal bath is possibly a prominent and experimentally pertinent one.

The experimental detection of the proposed non-unitary time dynamics of topological modes demands a controlled time modulation of the band gap through which topologically distinct insulating phases can be accessed across a quantum phase transition. Such a requirement can, in principle, be fulfilled in quantum crystals and cold atomic systems. In quantum materials external hydrostatic pressure, for example, can induce topological quantum phase transitions as has been demonstrated in pressured YbB$_6$~\cite{exp:0}. A time varying pressure thus can give rise to a time-dependent band gap in quantum materials, thereby offering a platform to test the non-unitary time dynamics of topological modes with the system-to-environment coupling possibly controllable in state-of-the-art experiments therein. On optical lattices a number of topological band engineering protocols has been proposed and realized~\cite{exp:1, exp:2, exp:3, exp:4}. On such systems, a time tunable band gap can be realized with a time-varying flavor-dependent (orbital and spin) on-site potential $m_0$. In cold atomic systems the system-to-environment coupling can be controlled efficiently by tuning the lattice potential, achievable with fluctuations in laser intensity and polarization, for instance. The proposed time evolution of the topological modes (melting and condensation) can be monitored by measuring the LDOS close to the zero-energy at various time instants. In quantum crystal such measurements can be performed with scanning tunneling microscope (STM)~\cite{STM:1, STM:2}. The LDOS in cold atomic systems can be measured by local radio frequency spectroscopy, for example~\cite{exp:5, exp:6, exp:7, exp:8}. The present work should therefore stimulate future theoretical and experimental investigations geared toward unfolding the time dynamics of topological modes in open quantum systems.

\acknowledgments

S.P.\ acknowledges support from the Kishore Vaigyanik Protsahan Yojana, Department of Science and Technology (DST), Government of India. We thank Sumilan Banerjee for insightful discussions and for providing access to the computational facility used in this work. B.R.\ was supported by NSF CAREER Grant No.\ DMR-2238679. We thank Vladimir Juri\v ci\' c and Christopher A.\ Leong for critical comments on the manuscript. Valuable discussions with Daniel Arovas, Ariel Sommer, and Sanjib Kumar Das are thankfully acknowledged.

\end{document}